\documentclass{aa}

\usepackage{natbib}
\usepackage{graphicx}
\usepackage[mathcal]{eucal}
\usepackage{amssymb}

\newcommand{\ds}{\displaystyle}
\newcommand{\inv} {\frac {1}}

\newcommand{\fig}[3]{
      \begin{figure}[ht]
        \begin{center}
        \resizebox{\hsize}{!}{\includegraphics{#1}}
        \end{center}
        \caption{#2}
        \label{#3}
        \end{figure} }
\newcommand{\eqn} [1] {
\begin{equation}#1
\end{equation}}
\newcommand{\eqna} [1] {
\begin{eqnarray}#1
\end{eqnarray}}
\newcommand{\latin}[1]{\it #1}

\newlength{\lenA} %
\newlength{\lenB} %
\begin{document}
\title{Excitation of solar-like oscillations across the HR diagram}
\author{Samadi R. \inst{1,2} \and Georgobiani D. \inst{3}
\and Trampedach R. \inst{4} \and Goupil M.J.\inst{2}
\and Stein R.F.  \inst{5} \and  Nordlund {\AA}.  \inst{6} }

\institute{
Observat\'orio Astron\'omico UC, Coimbra, Portugal \and
Observatoire de Paris, LESIA, CNRS UMR 8109, 92195 Meudon, France \and
Center for Turbulence Research, Stanford University NASA Ames Research Center, Moffett Field, USA \and
Research School of Astronomy and Astrophysics,
   Mt.\ Stromlo Observatory, Cotter Road, Weston ACT 2611, Australia \and
Department of Physics and Astronomy, Michigan State University, Lansing, USA \and
Niels Bohr Institute for Astronomy Physics and Geophysics, Copenhagen, Denmark
}
\offprints{R. Samadi}
\mail{Reza.Samadi@obspm.fr}
\date{\today} 

\titlerunning{Excitation of solar-like oscillations across the HR diagram}

\abstract
{}
{We extend   semi-analytical computations of excitation rates
 for solar oscillation modes to those of other solar-like oscillating stars  to compare them 
  with   recent observations}
{Numerical 3D~simulations of surface  convective zones of several solar-type
oscillating stars are used  to characterize the turbulent spectra as well as  to constrain  the convective
velocities and turbulent entropy fluctuations  in the uppermost part of the convective zone of such stars. 
These constraints, coupled with a theoretical model for stochastic excitation, provide
the rate ${\cal P}$ at which energy is injected into the p-modes by turbulent convection. 
These energy rates are compared with those derived directly from the 3D simulations. }
{The  excitation rates obtained from the 3D simulations are systematically lower than those computed from the semi-analytical excitation model. 
We find that ${\cal P}_{\rm max}$, the  ${\cal P}$  maximum, scales as $(L/M)^s$  where  $s$ is the slope of
the power law and  $L$ and $M$ are the mass and luminosity  of the 1D stellar model built consistently with the associated 3D~simulation.
The slope is found to  depend significantly on the adopted
form of $\chi_k$, the eddy time-correlation; using a Lorentzian, $\chi_k^{\rm L}$,
results in $s=2.6$, whereas a Gaussian, $\chi_k^{\rm G}$, gives $s=3.1$.
\\
Finally, values of $V_{\rm max}$, the maximum in the mode velocity, are estimated from
 the computed power laws for  ${\cal P}_{\rm max}$ and
we  find that  $V_{\max}$ increases  as $(L/M)^{sv}$.
Comparisons with  the currently available ground-based observations show that
the computations  assuming a Lorentzian $\chi_k$ yield  a slope, $sv$, closer to the observed one
than the slope obtained when assuming a Gaussian. We show that the
spatial resolution of  the 3D simulations  must be high enough to
obtain  accurate computed  energy rates.}
{}

 
\maketitle

\section{Introduction}

Stars with masses $M \lesssim 2 {\rm M}_\odot$ have upper
 convective zones where  stochastic excitation of  p-modes by  turbulent convection
takes place as in the case of
 the Sun. As such, these stars  are often  referred to as  {\it solar-like oscillating stars}.
 One of the major goals of the future space seismology mission CoRoT
 \citep{Baglin98}, is to measure the amplitudes and the line-widths of
 these  stochastically driven modes.  From the measurements of the
 mode line-widths  and amplitudes, it is possible to infer the rates at
 which acoustic modes are excited \citep[see {\protect{\latin{e.g.\ }}}][]{Baudin05}.
Such  measurements will then  provide valuable constraints on the theory of stellar oscillation  excitation and damping.
In turn, improved models of excitation and damping will provide valuable information
about convection in the outer layers of solar-like stars.

The mechanism of stochastic excitation has been modeled  by several authors
\citep[{\protect{\latin{e.g.\ }}}][ for a review see \citealt{Stein04}]{GK77,Osaki90,Balmforth92c,GMK94,Samadi00I}.
These models yield the  energy rate, ${\cal P}$, at which p-modes are excited by turbulent convection
but  require  an accurate knowledge of the time averaged and
 -- above all -- the \emph{dynamic}  properties  of turbulent convection.
\paragraph{Eddy time-correlations.} In the approach of \citet[][hereafter Paper\,I]{Samadi00I},
the \emph{dynamic}  properties of turbulent convection  are  represented by $\chi_k$,
the frequency component of the  auto-correlation product  of the turbulent velocity field;
$\chi_k$   can be   related to the convective eddy time-correlations.
 \citet[][hereafter Paper\,III]{Samadi02II} have shown that the Gaussian function
usually used for modeling $\chi_k$ is inappropriate and is at the origin of the
under-estimation of the computed maximum value of the solar p-modes excitation rates when compared with the observations.
On the other hand, the authors have shown that a Lorentzian profile provides the best fit to
the  frequency dependency of $\chi_k$ as  inferred from   a  3D simulation of the Sun.
Indeed, values  of ${\cal P}$ computed with the model of stochastic excitation of Paper\,I
and using a Lorentzian for $\chi_k=\chi_k^{\rm L}$
is better at reproducing the solar seismic observations
whereas a Gaussian function, $\chi_k^{\rm G}$, under-estimates the amplitudes of solar p-modes.
Provided that such a non-Gaussian model for  $\chi_k$  is assumed, the  model  of stochastic
 excitation is -- for the Sun -- rather satisfactory.
An open question, which we address in the present paper,   is whether such non-Gaussian behavior also stands for other
solar-like oscillating stars and what consequences arise for the theoretical excitation rates, ${\cal P}$.

\paragraph{Stochastic excitation in stars more luminous than the Sun.}
In the last five years, solar-like oscillations have been detected
in several stars (see for instance the review by \citet{Bedding03}).
Theoretical  calculations  result in an \emph{overestimation}  of their amplitudes \citep[see][]{Kjeldsen01,Houdek02}.
For instance, using Gough's \citeyearpar{Gough76,Gough77} non-local and time dependent
treatment of convection, \citet{Houdek99} have calculated
 expected values of $V_{\rm max}$, the  maximum oscillation amplitudes, for different  solar-like oscillating stars.
 Their calculations, based on a simplified excitation model, imply that
$V_{\rm max}$ of solar-type oscillations  scale
as $(L/M)^{1.5}$ where $L$ and $M$ are the luminosity and mass of the star \citep[see][ hereafter HG02 ]{Houdek02}.
A similar scaling law was empirically found earlier
 by \citet{Kjeldsen95}.
As pointed out by HG02, all these scaling laws overestimate
the observed amplitudes of solar-like oscillating stars
hotter and more massive than the Sun ({\latin{e.g.\ }}$\beta$\,Hydri, $\eta$\,Bootis, Procyon, $\xi$\,Hydrae).
As the mode amplitude results from a balance between excitation and damping, this
\emph{overestimation} of the mode amplitudes can be attributed either to an
overestimation of the excitation rates or an underestimation of the damping rates.
In turn, any overestimation of the excitation rates can be attributed
either to the excitation model \emph{itself} or to the underlying convection
model. 

All the related physical  processes are complex and  difficult
to model. The present excitation model therefore uses a number of approximations
such as the assumption of incompressibility, and the scale length separation between
the modes and the turbulent eddies exciting the modes.
 It has been shown that the current excitation model
 is valid in the case of the \emph{Sun}\ (Paper\,III), but
its validity in a broader region of the HR-diagram has not been confirmed until
now.

 Testing the validity of the theoretical model of stochastic excitation with the
help of 3D simulations of the outer layers of stellar models is the main goal of the present paper.
For that purpose, we compare the p-mode excitation rates for stars
with different temperatures and luminosities as obtained by  direct calculations and
by the semi-analytical method as outlined below.

Numerical 3D simulations enable one  to compute directly the excitation rates of p-modes for stars
with various temperatures and luminosities. For instance this was already undertaken for the Sun by
 \citet{Stein01II} using the numerical approach introduced in \citet{Stein01I}.
Such calculations  will next  be called ``{\bf direct} calculations''.
They are time-consuming and do not easily allow
massive  computations of the excitation rates for stars
with different temperatures and luminosities. On the other hand, an excitation model offers the advantage of
testing separately several properties entering the excitation mechanism
which are not well understood or modeled.  Furthermore, once it is validated,
it can be used  for  a large set of 1D models of stars.

As it was done for the Sun in \citet[][ hereafter Paper\,II]{Samadi02I}  and Paper\,III, 3D simulations can also  provide quantities which  can be implemented
in  a  formulation for the excitation rate ${\cal P}$, thus avoiding
 the use of the mixing-length approach with  the related free parameters,
and assumptions about the turbulent spectra.
Such calculations will next be called ``{\bf semi-analytical} calculations''.

We stress however that in any case, we cannot avoid the use of 1D~models for computing accurate
eigen-frequencies for the whole observed frequency range. In the present paper, the 1D models are 
constructed to be as consistent as possible with their corresponding 3D simulations,
as described in Sect. 3.

\medskip
 This paper is organized as follows:
In Sect.~2 we present the  methods considered here for computing ${\cal P}$, that is
the so-called ``direct'' method based on Nordlund \& Stein's \citeyearpar{Stein01I} approach (Sect.~2.1) and
 the so-called ``semi-analytical'' method based on the approach from Paper~I, with modifications
 as presented in Papers\,II \&  III and in the present paper (Sect.~2.2).

Comparisons between direct and semi-analytical calculations of the excitation rates
are performed in  seven  representative cases  of  solar-like oscillating stars.
 The  seven 3D simulations all  have the  same number of mesh points.
 Sect.~3 describes these simulations and their associated 1D stellar models.

The 3D simulations provide constraints on quantities related to the convective fluctuations,
in particular the eddy time-correlation function, $\chi_k$, which, as stressed above, plays an important role in the excitation of solar p-modes.
The  function $\chi_k$ is therefore inferred from each simulation
and compared with simple analytical function (Sect.~4).

Computations of the excitation rates of their associated p-modes are next undertaken in Sect.~5 using both the
direct approach and the  semi-analytical approach.
In the semi-analytical method, we employ model parameters as derived from the 3D simulations in Sect.~4.


In Sect.\ 5.2 we  derive the expected scaling laws for ${\cal P}_{\rm max}$, the maximum in ${\cal P}$,
as a function of $L/M$ with both  the direct  and semi-analytical methods
 and  compare the results.
This allows us to investigate the implications of such power laws for the expected values of $V_{\rm max}$
and to compare our results with the seismic observations of solar-like oscillations in Sect.\ 5.3. We also compare with previous 
theoretical results  \citep[{\protect{\latin{e.g.\ }}}][]{Kjeldsen95,Houdek02}.

We finally assess the validity of the present stochastic excitation model
and discuss the importance of the choice of the model for  $\chi_k$ in Sect.~6.

\section{Calculation of the p-mode excitation rates}
\label{sec:Calculation of the p mode excitation rates}

\subsection{The direct method}
  The energy input per unit time  into  a given stellar acoustic mode 
is calculated numerically according
  to Eq.~(74) of \citet{Stein01I}  multiplied by ${\cal S} $, the area of the simulation box, to get the excitation rate (in erg s$^{-1}$) :
\begin{equation}\label{dEdt}
{\cal P} (\omega_0) = \frac{\omega_0^2 \, {\cal S} }{8 \: \Delta \nu \: {\cal E}_{\omega_0}} \,
\left | \int_r dr \:  \Delta \hat P_{\rm nad} (r,\omega_0)\: { {\partial
\xi_{\rm r}} \over {\partial r} } \right |^2
\,
\end{equation}
where $\Delta \hat P_{\rm nad}(r,\omega)$ is the discrete Fourier component of the non-adiabatic
pressure fluctuations, $\Delta P_{\rm nad}(r,t)$,  estimated at the mode
eigenfrequency  
$\omega_0= 2 \pi \nu_0$, $\xi_{\rm r}$ is the radial component
of the  mode displacement eigenfunction,  $\Delta \nu=1/T$
the frequency resolution corresponding to the total simulation time
$T$ and  ${\ds {\cal E}_{\omega_0}}$  is the mode energy per unit
surface area defined  in \citet[][ their Eq.~(63)]{Stein01I} as:
\begin{equation}
{\cal E}_{\omega_0}= { 1 \over 2 } \, \omega_0^2 \, \int_r dr \: \xi_{\rm r}^2 \, \rho \left ( { r \over R } \right )^2 \; .
\end{equation}
 Note that Eq.\ (\ref{dEdt}) corresponds to the direct calculation of $P dV$ work of the non-adiabatic gas and turbulent pressure (entropy and Reynolds stress) fluctuations on the modes.   
The energy in the denominator of Eq.\,(\ref{dEdt}) is essentially the mode mass.  The additional
factor which turns it into energy is the mode squared amplitude which is arbitrary
and cancels the mode  squared amplitude in the numerator.  For a given driving (i.e. $P\, dV$ work), the variation of the mode energy is inversely proportional to the mode energy  \citep[see Sect. 3.2 of][]{Stein01I}. Hence, for a given driving, the larger the mode energy 
({\it i.e.}, the mode mass or mode inertia) the smaller the excitation rate.  

 In Eq.\ (1) the  non-adiabatic Lagrangian pressure fluctuation,
  $\Delta \hat  P_{\rm nad} (r,\omega)$, is calculated as the following: 
We first compute the non-adiabatic  pressure fluctuations
$\Delta P_{\rm nad}(r,t)$  according to Eq.\,\ref{dpnad_2} in
Appendix A. 
We then  perform the temporal Fourier transform of
$\Delta P_{\rm nad}(r,t) $ at each depth $r$ to get
$\Delta \hat P_{\rm nad} (r,\omega)$. 

The mode displacement eigenfunction
$\xi_{\rm r} (r)$ and the mode eigenfrequency $\omega_0$  are
calculated as explained in Section~3. 
Its vertical derivative, ${\partial \xi_{\rm r}} / {\partial r}$, is
normalized by  the mode energy per unit surface area, ${\ds {\cal
    E}_{\omega_0}}$, and then multiplied by $\Delta \hat P_{\rm nad}$.
The result is integrated over
the simulation depth, squared and divided by $8\, \Delta \nu$.
We next multiply the result by the area of the simulation box ($\cal S$) to obtain ${\cal P}$, the total excitation rates in erg s$^{-1}$ for the entire star.
Indeed the nonadiabatic pressure fluctuations are uncorrelated on
large scales, so that average $\Delta P_{\rm nad} ^2$ is inversely
proportional to the area. Multiplication by the area of the 
stellar simulation gives the excitation rates for the entire star
as long as the domain size is sufficiently large to include
several granules.

\subsection{The semi-analytical method}
 \label{sec:The semi-analytical method}

Calculations of excitation rates by the semi-analytical method are
based on a  model of stochastic excitation. The excitation model we consider is the same as presented in Paper~I.
In this model of excitation and in contrast to previous models \citep[e.g.][]{GK77,Balmforth92c,GMK94},
the driving by turbulent convection is ensured not only by
the Reynolds stress tensor but also by the advection of the turbulent fluctuations of entropy
by the turbulent movements (the so-called entropy source term).

As in Paper~I,  we consider only radial p-modes.
 We do not expect any significant differences for low $\ell$ degree modes.
 Indeed, in the region where the excitation takes place,  the low  $\ell$ degree modes have the same behavior
 as the radial modes. Only for very high $\ell$ degree modes ($\ell \gg 100$) -~ which will not be seen
 in stars other than the Sun~- can a significant effect be expected, as is quantitatively confirmed
 (work in progress).

The excitation rates are computed  as in Paper~II, except for the change
detailed below.
The rate at which a given mode with frequency $\omega_0=2\pi \nu_0$ is excited
is then calculated with the set of Eqs.\ (1)--(11) of Paper~II.
These equations are based on the excitation model of Paper~I, but assume that injection of acoustic energy
into the modes is isotropic.
However, Eq.\ (10) of Paper~II must be corrected for an analytical error
  \citep[see ][]{Samadi05c}. This  yields the following correct expression for Eq.\ (10) of Paper~II:
\eqna{
S_{R}(r,\omega_0) & = & \int_0^\infty { {\rm d} k \over  k^2} \,
  \frac{E(k,r)}{u_0^2} \, \frac{E(k,r)}{u_0^2} \, \nonumber \\ & & \times \int_{-\infty}^{+\infty} {\rm d}\omega \, \chi_k ( \omega_0 + \omega, r) \, \chi_k (\omega, r) \label{eqn:SR}
}
where $u_0=\sqrt{\Phi/3} \, \bar{u}$, $\Phi$ is Gough's \citeyearpar{Gough77} anisotropy factor, $\bar{u}$  is the rms value of  $\vec u$, the  turbulent velocity field, $k$ the wavenumber  and $\chi_k(\omega)$
is  the frequency component of the  correlation product  of $\vec u$.

The method then requires the knowledge of  a number of  input parameters which are of three different types:
\begin{itemize}
\item[1)] Quantities which are related to the oscillation modes:
the eigenfunctions ($\xi_r$) and associated eigen-frequencies ($\omega_0$).
\item[2)] Quantities which are related to the spatial and time averaged properties of the medium:
the mean density ($\rho_0$), $\alpha_s \equiv \langle\left (\partial p / \partial s \right )_\rho\rangle $ --
where $s$ is the entropy, $p$ the gas pressure and $\langle\dots\rangle$ denotes horizontal and time averages --
the mean square of the vertical component of the  convective velocity, $\langle w^2\rangle$,
the mean square of the entropy fluctuations, $\langle\tilde s^2\rangle$, and the mean anisotropy, $\Phi$ (Eq.\ (2) of Paper~II).
\item[3)] Quantities which contain information about spatial and temporal auto-correlations of the convective fluctuations:
the spatial spectrum of
the turbulent kinetic energy and entropy fluctuations, $E(k)$ and $E_s(k)$,
respectively, as well as
the temporal spectrum of the correlation product of the turbulent velocity
field, $\chi_k$.
\end{itemize}

Eigen-frequencies and  eigenfunctions [in 1) above] are computed with the
adiabatic pulsation code ADIPLS \citep{JCD91b} for each of the 1D~models associated with the 3D simulations (see  Sect.\ \ref{sec:The simulated stars and their associated 1D models}).

The spatial and time averaged quantities (in 2) and 3) above)
are obtained from the 3D simulations in the  manner of Paper~II.
For $E(k)$, however, we use the actual spectrum as calculated from
the 3D simulations and not an analytical fit as was done in Paper~II.
However as in Paper~II, we assume for  $E_s(k)$  the  $k$-dependency of $E(k)$
(we have checked  this assumption for one simulation and found no significant change in ${\cal P}$).

For each simulation, we determine  $\chi_k$ as in Paper~III ({\latin{cf.\ }}Sect.\ \ref{Inferred properties of  turbulent convection}).
Each $\chi_k$ is then compared  with various analytical forms, among which some were investigated in Paper~III.
Finally  we select the analytical forms which are  the closest to the behavior of  $\chi_k$
 and use them, in Section~\ref{sec:p-mode excitation rates accross the HR diagram},
to compute ${\cal P}$.

\section{The convection simulations and their associated 1D models}
\label{sec:The simulated stars and their associated 1D models}

Numerical simulations of surface convection for seven different
solar-like stars were performed by \citet{Trampedach99b}.
\begin{figure}[ht]
 \resizebox{\hsize}{!}{\includegraphics{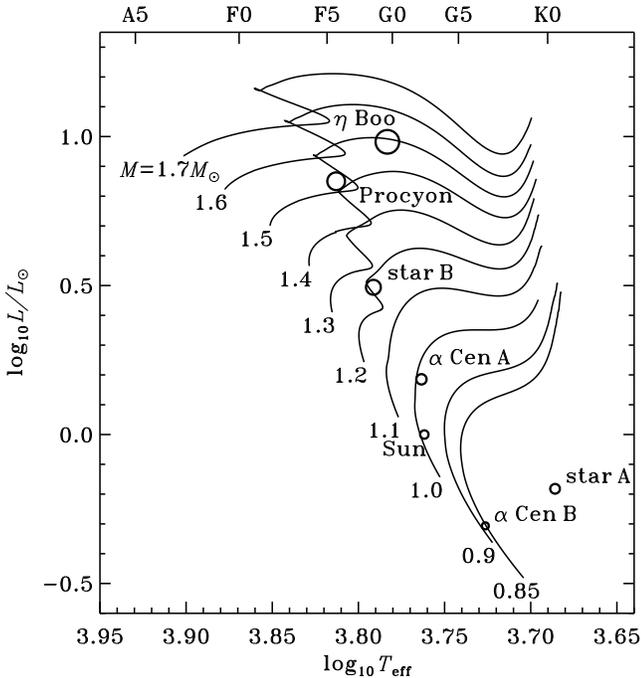}} 
\caption{Location of the convection simulations in the HR
diagram. The symbol sizes vary proportionally to the stellar
radii. Evolutionary tracks of stars, with masses as indicated, were calculated on the base of Christensen-Dalsgaard's stellar evolutionary code \citep{JCD82,JCD83a}.
\label{HR7}}
\end{figure}
These hydrodynamical simulations are characterized by the effective
temperature, $T_{\rm eff}$ and acceleration of gravity, $g$,
as listed in Table \ref{tab:simus}.
The solar simulation with
the same input physics and  number of mesh points is included for
comparison purposes. The surface gravity is an input parameter,
while the effective temperature is adjusted by changing the
entropy of the inflowing gas at the bottom boundary.
The simulations have 50$\times$50$\times$82
grid points. All of the models have solar-like chemical composition,
with hydrogen abundance $X=0.703$ and metal abundance $Z=0.0145$.
The simulation time-series all cover at least five periods of the primary
p-modes (highest amplitude, one node at the bottom boundary),
and as such should be sufficiently long. 


The convection simulations are shallow (only a few percent of the stellar
radius) and therefore contain
only few modes. To obtain mode eigenfunctions, the simulated
domains are augmented by 1D envelope models in the interior by
means of the stellar envelope code by \citet{JCD83a}. Convection in the
envelope models is based on the mixing-length formalism \citep{Bohm58}.

\citet{trampedach:alfa-fit} fit 1D stellar envelopes to
average stratifications of the seven convection simulations
by matching temperature
and density at a common pressure point near the bottom of the simulations.
The fitting parameters are the mixing-length parameter, $\alpha$, and a
form-factor, $\beta$, in the expression for turbulent pressure:
$P_{\rm turb}^{\rm 1D} = \beta\varrho u_{\rm MLT}^2$,
where $u_{\rm MLT}$ is the
convective velocities predicted by the mixing-length formulation.
A consistent matching of the simulations and 1D envelopes is
achieved by using the same   equation of state (EOS) by
\citet[][also referred to as the MHD EOS, with reference to Mihalas, Hummer, and D\"appen]{Werner88}
and opacity distribution functions (ODF) by
\citet{Kurucz92a,Kurucz92b}, and also by using
$T$-$\tau$ relations derived
from the simulations \citep{trampedach:T-tau}.

The average stratifications of the 3D simulations, augmented by the
fitted 1D envelope models in the interior, were used as the basis for
the eigenmode calculations using the adiabatic pulsation code by
\citet{JCD91b}. These combinations of averaged 3D simulations and matched
1D envelope models will, from hereon, be referred to as the 1D models.

The positions of the models in the HR diagram are presented in Figure\ \ref{HR7} and their
global parameters are listed in Table \ref{tab:models}.
Five of the seven models   correspond to actual stars, while
 Star\,A and Star\,B are merely sets of atmospheric parameters;
their masses and luminosities are therefore assigned somewhat arbitrarily
(the $L/M$-ratios, only depending on $T_{\rm eff}$ and $g$, are of course not
arbitrary).

\begin{table*}[ht]
\begin{center}
\begin{tabular}{lrcccccccc}
star  & $t_{\rm sim}$   & size &  $\log g$ & $T_{\rm eff}$ & $H_p$ & $L_{\rm h}/H_p$ & $C_s$ &  $t_s$& $t_{\rm sim}/t_{\rm s}$\\
&      [min] & [Mm$^3$] & & [K] & [km] & &  [km s$^{-1}$]  & [s] & \\
\hline
$\alpha$\,Cen\,B &  59~~~ &  4.0$\times$~4.0$\times$~2.2  &  4.5568 & 5363 & 95. & 42.1 &  7.49 & 12.72 & 278.3 \\
Sun              &  96~~~ &  6.0$\times$~6.0$\times$~3.4  &  4.4377 & 5802 & 134 & 44.8 &  7.78 & 17.30 & 332.9  \\
Star\,A          &  80~~~ & 11.6$\times$11.6$\times$~6.4  &  4.0946 & 4851 & 316 & 36.7 &  7.98 & 39.66 & 121.0\\
$\alpha$\,Cen\,A &  44~~~ &  8.9$\times$~8.8$\times$~5.1  &  4.2946 & 5768 & 189 & 47.1 &  7.81 & 24.17 & 109.2\\
Star\,B          & 110~~~ & 20.7$\times$20.7$\times$11.3  &  4.0350 & 6167 & 359 & 57.7 &  7.76 & 46.29 & 142.6\\
Procyon          & 119~~~ & 20.7$\times$20.7$\times$10.9  &  4.0350 & 6470 & 380 & 54.5 &  7.52 & 50.50 & 141.4 \\
$\eta$\,Boo      & 141~~~ & 36.9$\times$36.9$\times$16.3  &  3.7534 & 6023 & 709 & 52.0 &  7.40 & 96.13 & 88.0
\end{tabular}
\end{center}
\caption{Characteristics of the convection 3D simulations:  $t_{\rm sim}$ is the duration of the
relaxed simulations used in the present analysis, $H_p$ is the pressure scale height at the surface, $L_{\rm h}$ the size of the box in the horizontal direction,  $C_s$ the sound speed and  $t_{\rm s}$ the sound travel time across  $H_p$. All the simulations
have a spatial  grid of 50$\times$50$\times$82. }
\label{tab:simus}
\end{table*}

\begin{table}[ht]
\begin{center}
\begin{tabular}{lccccc}
  star &  $T_{\rm eff}$ & $M/{\rm M}_\odot$ & $R/R_\odot$ &$L/L_\odot$ & $L M_\odot/M L_\odot$ \\
       & [K]   & & & &\\
\hline
$\alpha$\,Cen\,B &  5363 & 0.90 & 0.827 & 0.51 & 0.56 \\
Sun              &  5802 & 1.00 & 1.000 & 1.02 & 1.02\\
Star\,A          &  4852 & 0.60 & 1.150 & 0.66 & 1.10\\
$\alpha$\,Cen\,A &  5768 & 1.08 & 1.228 & 1.50 & 1.38 \\
Star\,B          &  6167 & 1.24 & 1.769 & 4.07 & 3.28\\
Procyon          &  6470 & 1.75 & 2.102 & 6.96 & 3.98  \\
$\eta$\,Boo      &  6023 & 1.63 & 2.805 & 9.31 & 5.71
\end{tabular}
\end{center}
\caption{Fundamental parameters of the 1D-models associated with the 3D simulations of Table~\ref{tab:simus} }
\label{tab:models}
\end{table}

\section{Inferred properties of $\chi_k$ }
\label{Inferred properties of  turbulent convection}

\begin{figure}[ht]
 \begin{center}
\includegraphics[width=\lenA] {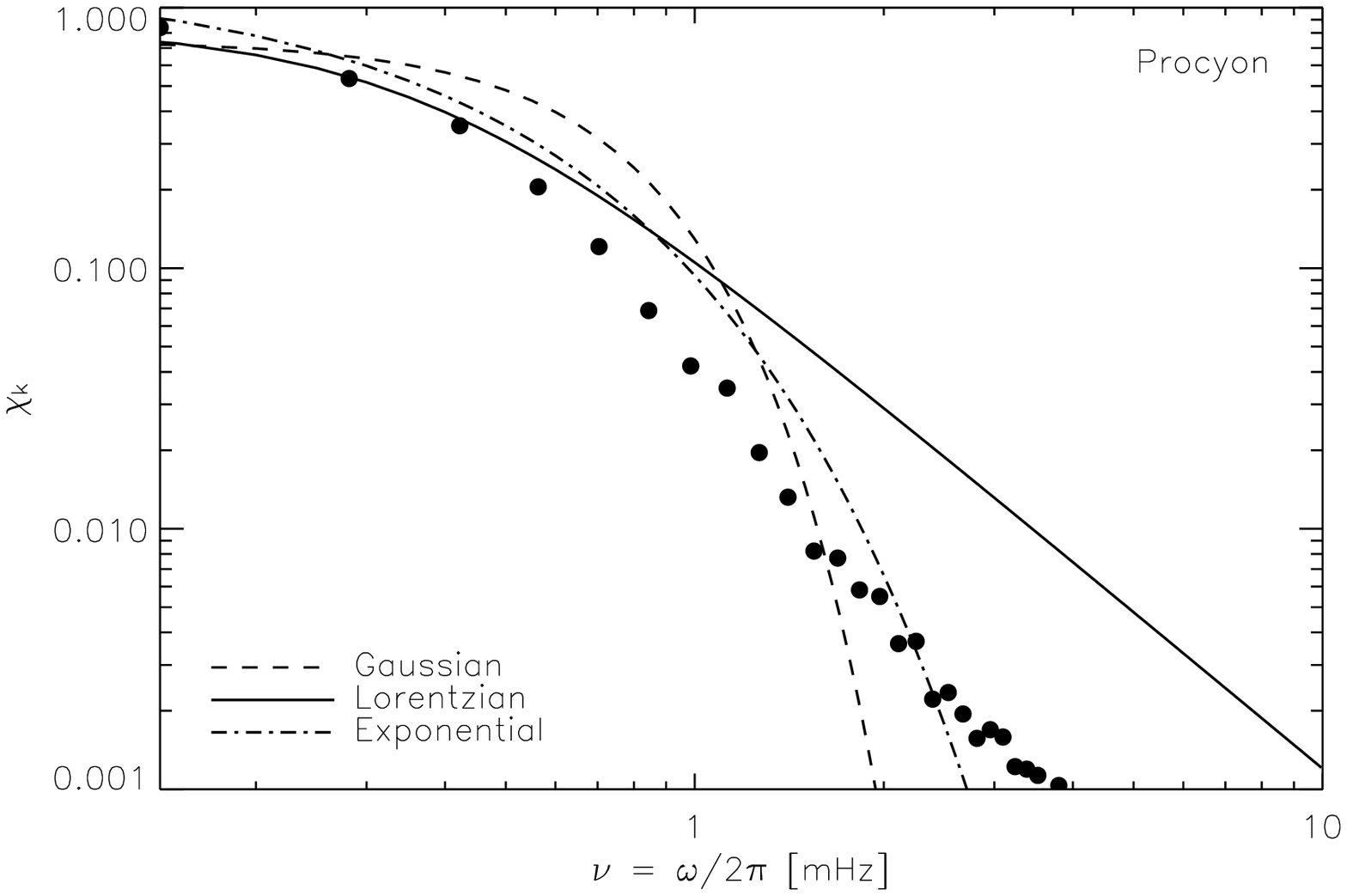} 
\includegraphics[width=\lenA] {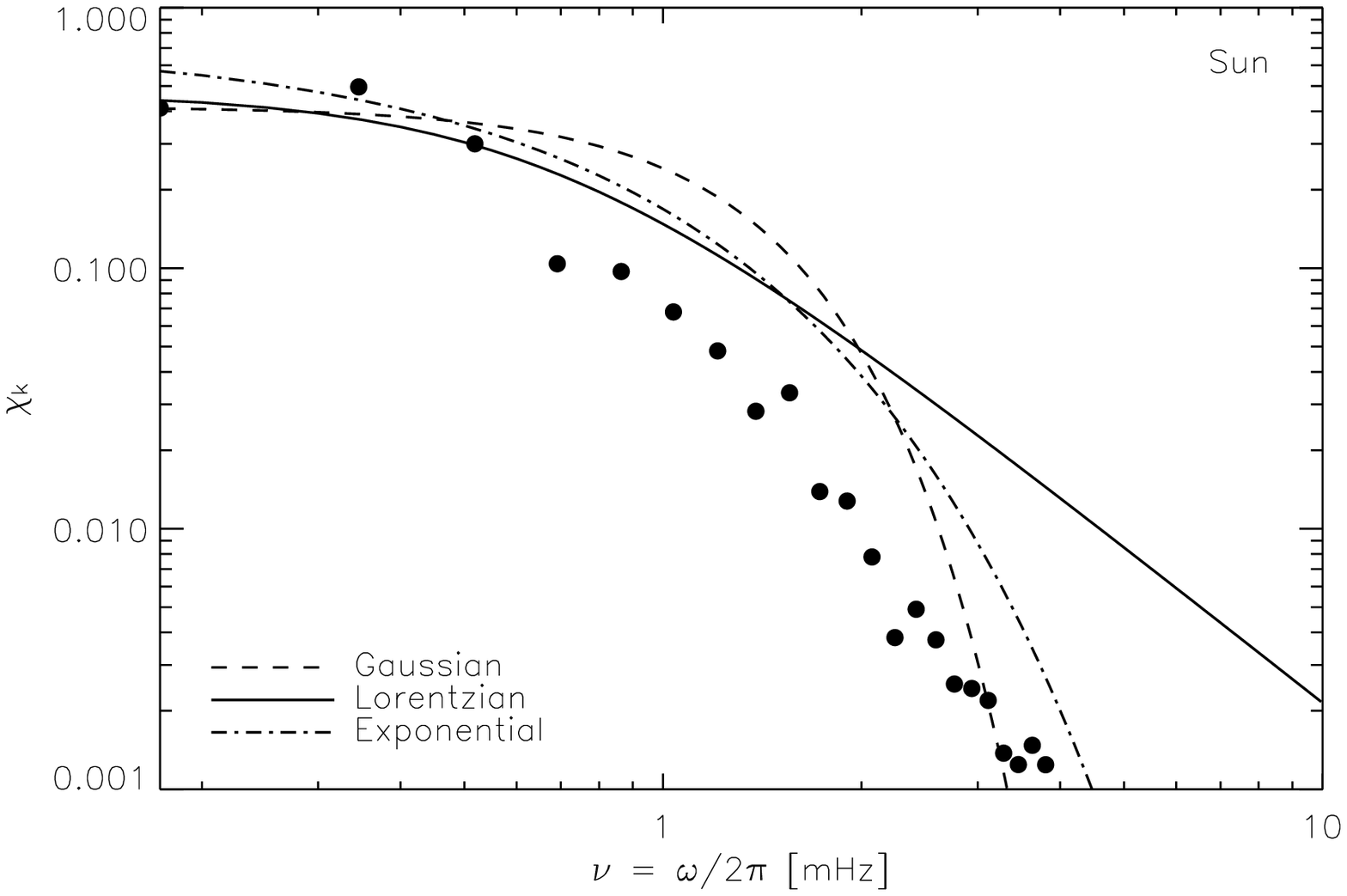}  
\includegraphics[width=\lenA] {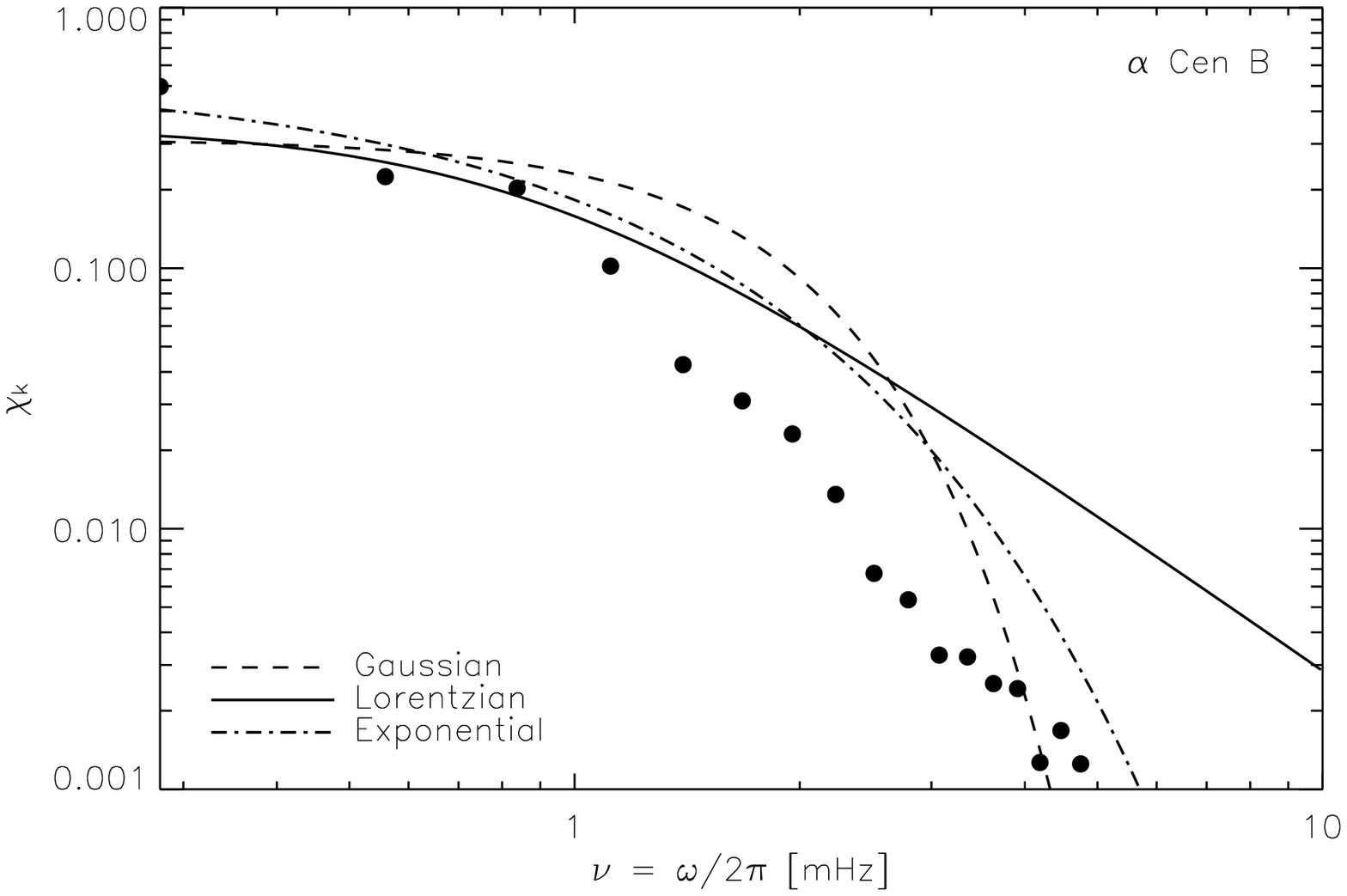} 
         \end{center}
\caption{The filled dots represent  $\chi_k$ obtained from the 3D simulations
for the wavenumber $k$ at which $E(k)$ is maximum and at the layer where  the excitation is maximum in the simulation.
The results  are presented  for three simulations:  Procyon (top), the Sun (middle) and $\alpha$\,Cen\,B (bottom).
The solid curves represent the Lorentzian form, Eq.\ (\ref{eqn:LF}), the dashed curves
the Gaussian form Eq.\ (\ref{eqn:GF}), and the dot dashed curves the exponential form Eq.\ (\ref{eqn:EF}). }
        \label{fig:kwpower}
\end{figure}

\begin{figure}[ht]
 \begin{center}
\includegraphics[width=\lenA] {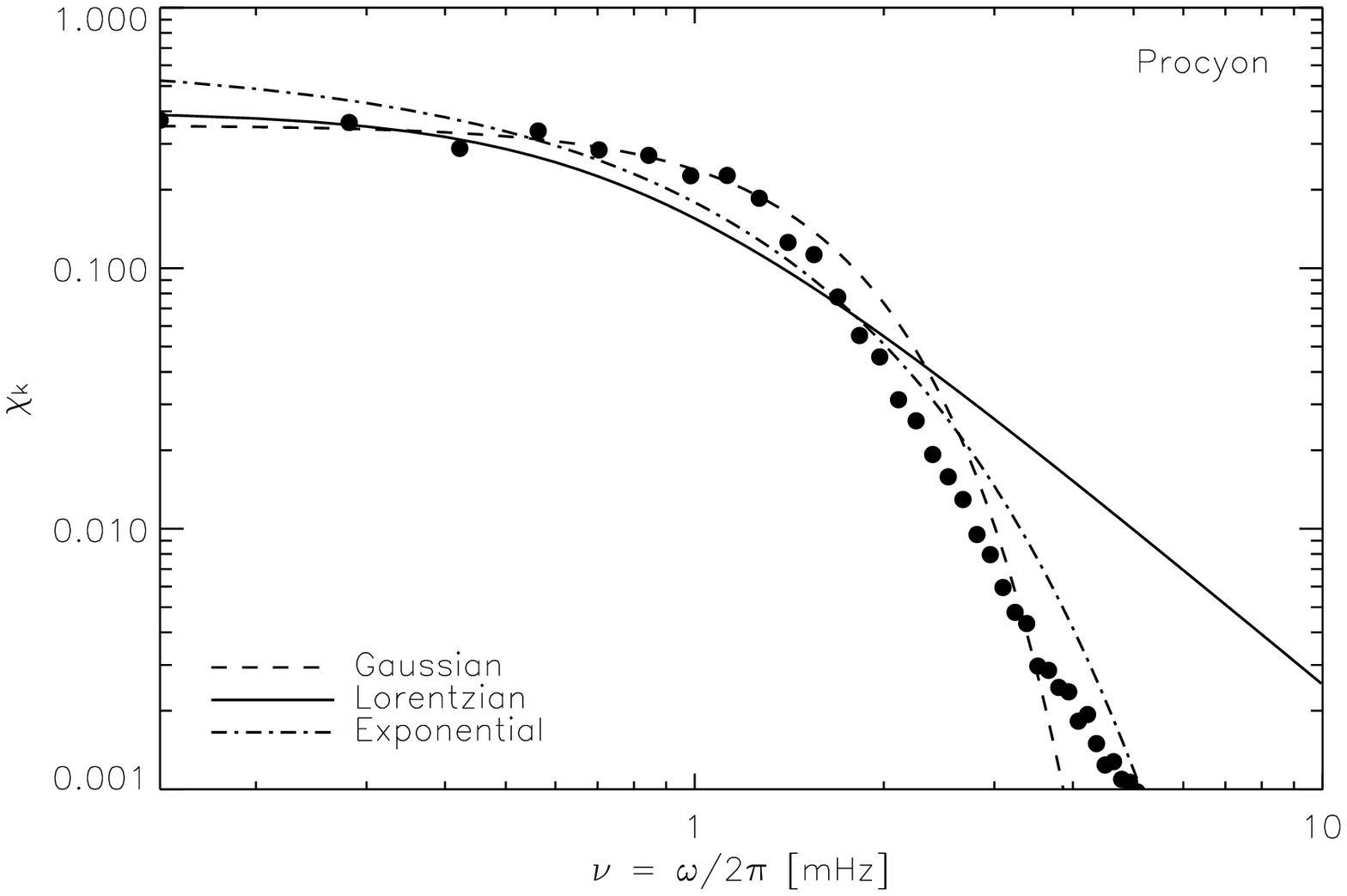} 
\includegraphics[width=\lenA] {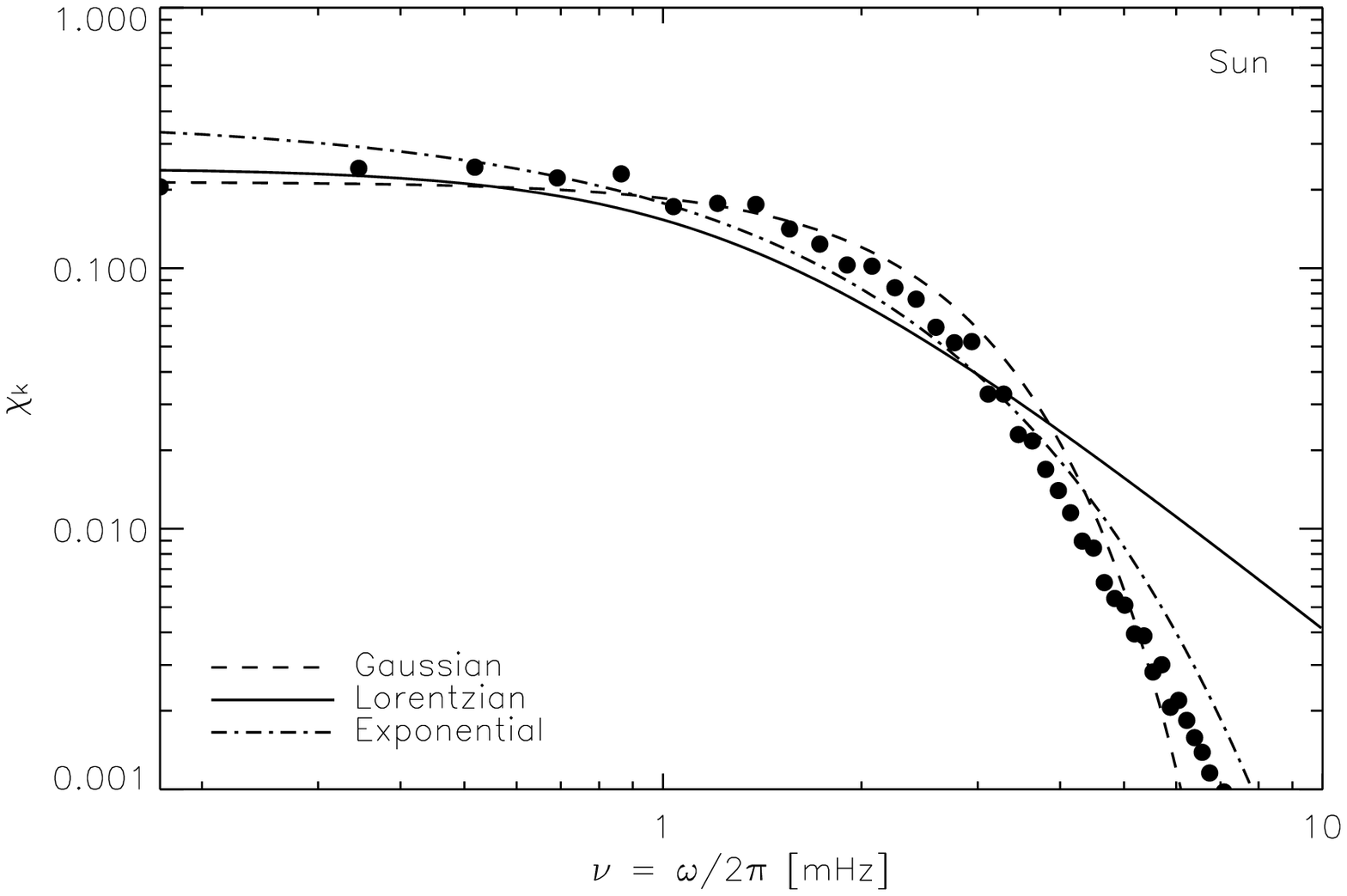} 
\includegraphics[width=\lenA] {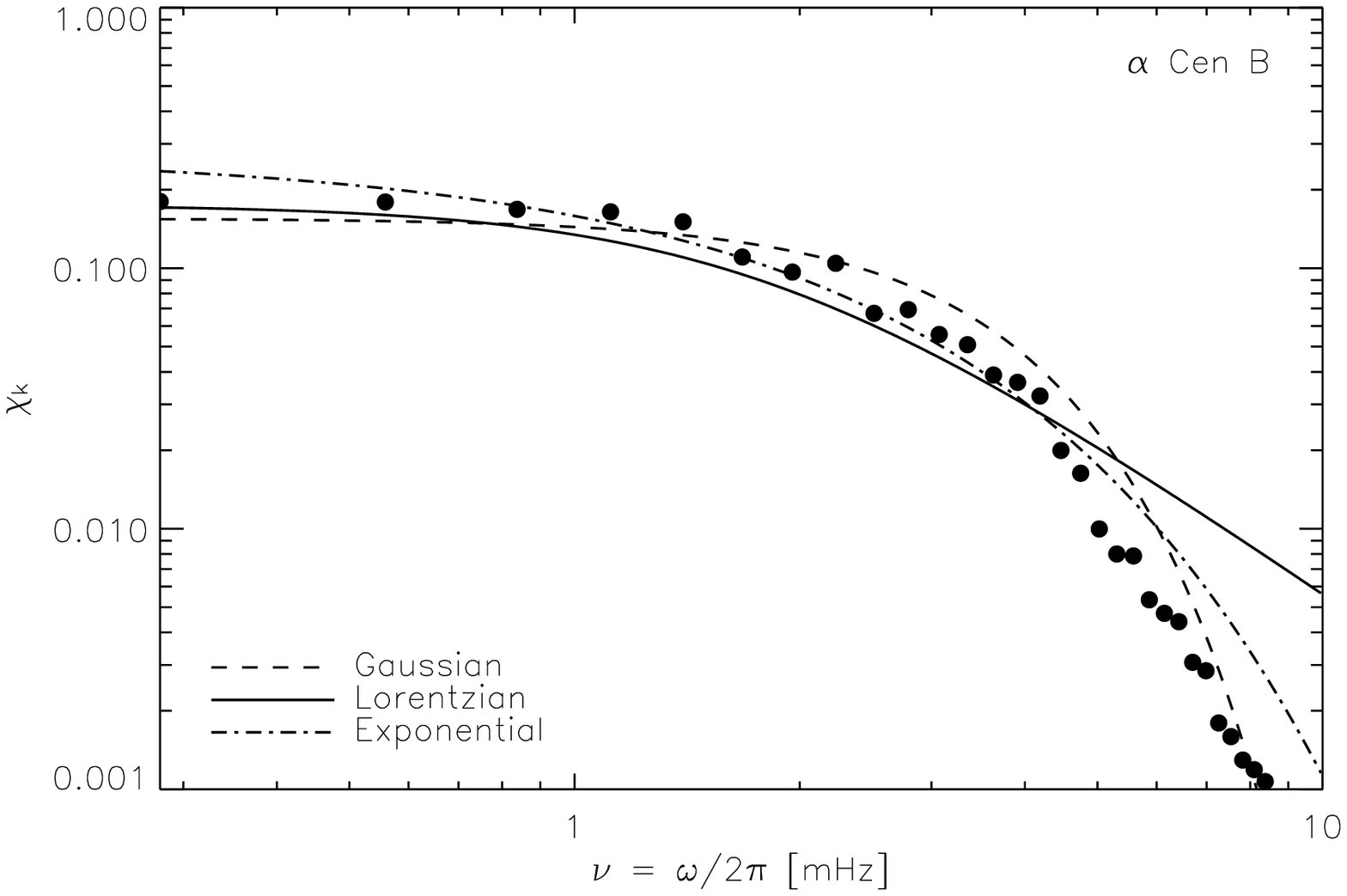} 
         \end{center}
\caption{Same as Fig.\ \ref{fig:kwpower}
for  $k=10~ k_{\rm min}$ where $k_{\rm min}$ is the first non-zero  wavenumber of the  simulation.}
        \label{fig:kwpower:K10}
\end{figure}

For each simulation, $\chi_k(\omega)$ is computed over the whole wavenumber ($k$) range covered by the simulations
and at different layers within the region where modes are excited.
We present  the results  at the layer  where the excitation is maximum,
{\latin{i.e.}}, where $u_0$ is maximum, and for two representative wavenumbers:
$k=k_{\max}$ at which $E(k)$ peaks and $k=10\,k_{\min}$,
where $k_{\min}$ is the first non-zero  wavenumber of the simulations.
Indeed, the amount of acoustic energy going into a given mode is largest
at this layer and at the  wavenumber  $k \simeq k_{\max}$,
provided that the mode frequency satisfies:  $\omega_0 \lesssim  (k_{\max} \, u_0)$.
Above   $ \omega_0 \sim  k_{\max} \, u_0  $, the efficiency of the excitation
decreases rapidly. Therefore low and  intermediate frequency  modes
({\latin{i.e.}},  $\omega_0 \lesssim  k_{\max} \, u_0$) are predominantly excited
at  $k \simeq k_{max}$. On the  other hand, high frequency modes are
predominantly excited by small-scale fluctuations, {\latin{i.e.}} at large $k$.
The exact choice of the  representative large wavenumber is  quite arbitrary;
 however it cannot be too large because of the  limited  number of mesh points $k \lesssim 25 ~k_{\rm min}$
and in any case, the excitation is negligible above  $k \simeq 20~k_{\rm min}$.
We thus chose the intermediate wavenumber  $k=10~k_{\rm min}$.
Fig.\ \ref{fig:kwpower} presents $\chi_k$ as obtained from the 3D simulations of Procyon,
$\alpha$\,Cen\,B and the Sun, at the layer where $u_0$ is maximum
and for the wavenumber $k_{\rm max}$.
Although defined as a function of $\omega$, for convenience,
 $\chi_k$  is plotted as a function of $\nu= \omega/2\pi$ throughout this paper.
Fig.\ \ref{fig:kwpower:K10} displays  $\chi_k$ for $k=10~k_{\rm min}$.
Results for the other simulations are not shown, as the results for Procyon,
$\alpha$\,Cen\,B and the Sun correspond to three representative cases.

%
%

In practice, it is not easy to implement directly in the excitation model the $\nu$-variation of  $\chi_k$ inferred from the 3D
simulations. An alternative and convenient way to compute ${\cal P}$
is to use simple analytical functions for  $\chi_k$ which are chosen so as to best represent the 3D results.
We then compare   $\chi_k$ computed with the 3D simulations with   the following simple analytical forms:
the Gaussian form
\eqn{
\chi_k^{\rm G}(\omega ) = \inv  { \omega_k \, \sqrt{\pi}}  e^{-(\omega / \omega_k)^2} \;,
\label{eqn:GF}
}
the Lorentzian form
\eqn{
\chi_k^{\rm L} (\omega ) =\frac{1} {\pi \omega_k/2} \,\frac{1}{1+ \left( 2 \omega / \omega_k \right )^2} \;,
\label{eqn:LF}
}
and the exponential form
\eqn{
\chi_k^{\rm E} (\omega ) =
\frac{1}{\omega_k}  e ^{-| 2 \omega/\omega_k | } \;.
\label{eqn:EF}
}
In Eqs.(\ref{eqn:GF}-\ref{eqn:EF}),  $\omega_k$ is the  line-width of the analytical function
 and is  related to  the velocity  $u_k$ of the eddy with wave number $k$ as:
\eqn{
\omega_k \equiv 2 \, {k u_k } \label{eqn:omegak}
}
In Eq.\ (\ref{eqn:omegak}),  $u_k$ is calculated from the kinetic  energy spectrum $E(k)$  as  \citep{Stein67}
\eqn{
u_k^2 =  \int_k^{2 k}  dk \, E(k)
\label{eqn:uk2}
}

As shown in Fig.\ \ref{fig:kwpower} and Fig.\ \ref{fig:kwpower:K10},
the Lorentzian $\chi_k^{\rm L}$ does not reproduce the $\nu$-variation of
$\chi_k$ satisfactorily. This is particularly true for the solar  case.
This contrast with the results of Paper~III where  it was found that $\chi_k^{\rm L}$ reproduces
nicely -- at the wavenumber where $E$ is maximum --
the $\nu$-variation of  $\chi_k$  inferred from the solar simulation investigated in  paper~III.
These differences in the results for the solar case can be explained by the {\emph low} spatial resolution of the present solar  simulation compared with that of Paper~III.
Indeed we have compared different solar simulations with different spatial resolution and
 found that the   $\nu$-variation of  $\chi_k$ converges to that of $\chi_k^{\rm L}$ as the spatial resolution increases (not shown here).
This dependency of $\chi_k$ with  spatial resolution of the simulation is likely to hold
for the non-solar simulations as well.   This result then suggests that
$\chi_k$ is in fact best represented by the Lorentzian form, $\chi_k^{\rm L}$.

As a consequence, realistic  excitation rates evaluated directly for a
convection simulation should be based on simulations with higher spatial resolution.
However the main goal of the present work is to test the excitation model, which can be done with the present set of simulations.
Indeed, we only need to use as inputs
for the excitation model the quantities related to the turbulent  convection ($E(k)$, $\chi_k$,\dots) as they  are in the simulations,
no matter how the real properties of  $\chi_k$ are.

For the present set of simulations,  we compare three analytical forms of $\chi_k$: Lorentzian, Gaussian and exponential.
For large $k$,  $\chi_k$  is overall  best modeled by a Gaussian (see Fig.\ \ref{fig:kwpower:K10} for $k=10~k_{\rm min}$).
For small $k$ (see Fig.\ \ref{fig:kwpower} for $k=k_{\rm max}$) both the exponential and the Gaussian are closer to $\chi_k$ than the Lorentzian.

For a given simulation,
depending  on the frequency, differences between $\chi_k(\nu)$ and the analytical forms are more or less pronounced.
%
%

The discrepancy between  $\chi_k(\nu)$ inferred from the 3D simulations and the exponential or the Gaussian forms
vary systematically with stellar parameters; decreasing as the convection
gets more forceful, as measured by, {\latin{e.g.}}, the turbulent- to total-pressure
ratio. Of the three simulations illustrated in Fig.\ \ref{fig:kwpower}, Procyon
has the largest and $\alpha$\,Cen\,B has the smallest
$P_{\rm turb}/P_{\rm tot}$-ratio.

As a whole for the different simulations and scale lengths $k$, we conclude that
the $\nu$-variation  of $\chi_k$ in the present set of simulations lies between
that of a Gaussian and an exponential.  However, neither of them is completely satisfactory. Actually a recent detailed study by \citet[][ in preparation]{Georgobiani04} tends to show that $\chi_k$ cannot
systematically be represented at all wavenumbers by a simple form such as a Gaussian, an exponential or a Lorentzian, but rather needs a more generalized power law. 
Hence, more sophisticated fits closer to the simulated $\nu$-variation of  $\chi_k$ could have been
considered, but for the sake of simplicity we chose to limit ourselves to the
three forms presented here.

\section{p-mode excitation rates across the HR diagram}
\label{sec:p-mode excitation rates accross the HR diagram}
\subsection{Excitation rate spectra (${\cal P}(\nu)$)}

For each simulation, the rates ${\cal P}$ at which the p-modes of the associated
1D~models are excited are computed both directly from the 3D simulations
and with the semi-analytical method (see Sect.\ \ref{sec:Calculation of the p mode excitation rates}).
In this section, the semi-analytical calculations are based on two  analytical forms of
$\chi_k$:  a Gaussian and  an exponential form as described in Sect.4.
The Lorentzian form as introduced in Sect.\ \ref{Inferred properties of  turbulent convection} is not investigated in the present section. Indeed our purpose here is to test the model of stochastic excitation by using
  constraints from the 3D simulations, and a Lorentzian behaviour is never obtained in the present
   3D simulations.

\begin{figure*}[h!]
        \begin{center}
\includegraphics[width=15.truecm]{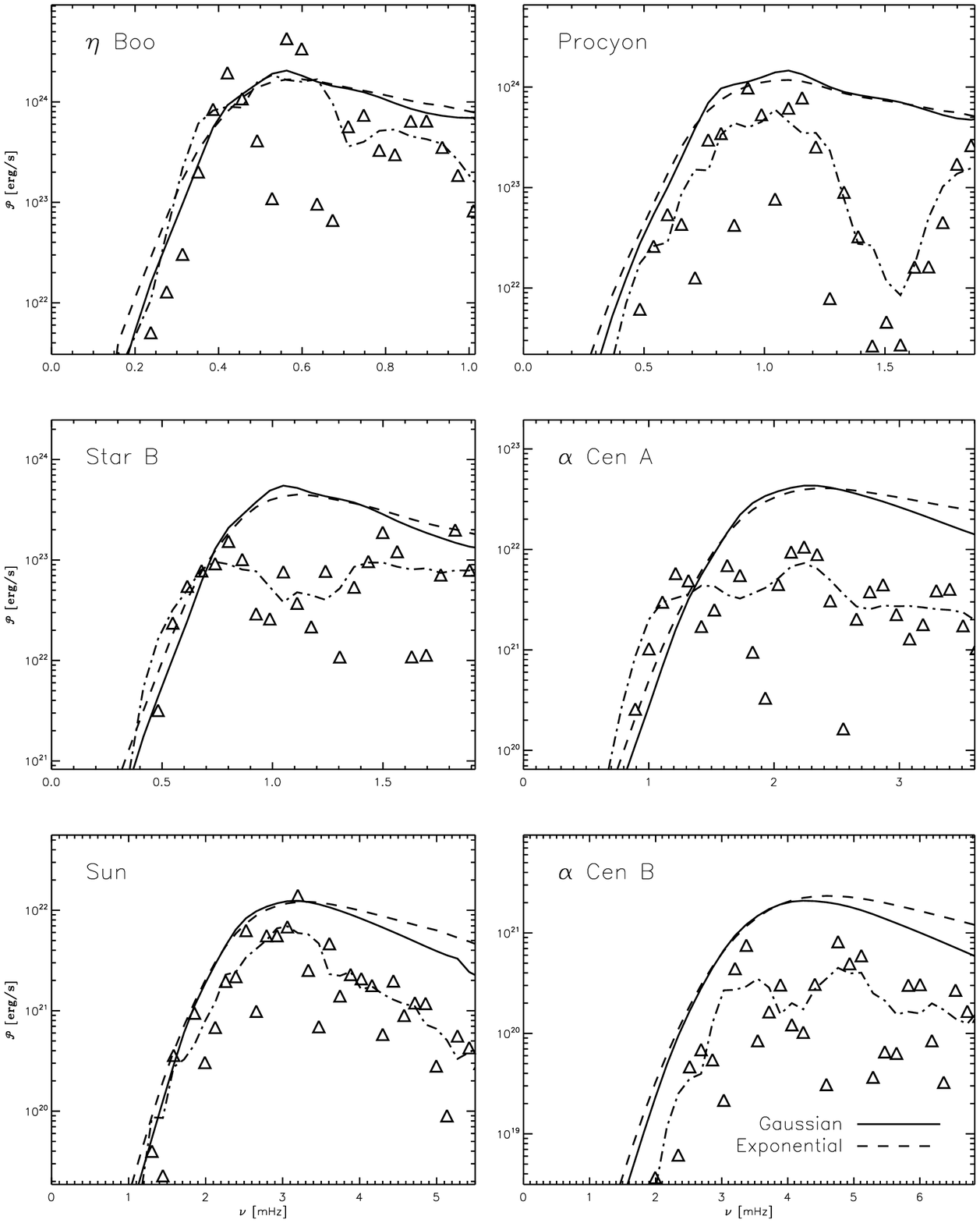} 
        \end{center}
\caption{Excitation rates, ${\cal P}$, are presented as functions of mode frequency
for six of the seven convection simulations listed in Tables \ref{tab:simus} and
\ref{tab:models}.
Each triangle corresponds to a single evaluation of the 3D work
integral estimated for a given eigenfrequency  according to
Eq.~(\ref{dEdt}).
 The dot-dashed lines correspond to a running mean of the triangle
  symbols performed over five frequencies.
The solid and dashed lines correspond to the excitation rates
calculated with the semi-analytical method and using the Gaussian and the
exponential forms of $\chi_k$, respectively. 
All results shown are obtained as  the sum of contributions from
the two sources of excitation:  excitation by the turbulent pressure
and  excitation by the  non-adiabatic gas pressure.
}
\label{fig:P}
   \end{figure*}

The results of the calculations of ${\cal P}$ using both  methods
are presented in Fig.\ \ref{fig:P} for the six most representative
simulations.
 In order to remove the large scattering in the direct
  calculations, we perform a  running
  mean over five frequency bins. The results of this averaging are shown
  by dot-dashed lines. The choice of five frequency bins is somewhat
  arbitrary. However we notice that between 2 to 10  frequency bins, the
  maximum and the shape of the spectrum do not significantly  change.  

 Comparisons  between  direct and  semi-analytical calculations using either $\chi_k^{\rm G}$
or $\chi_k^{\rm E}$  all show systematic differences: the excitation rates obtained with the direct calculations 
are systematically lower than those resulting from  the semi-analytical method. 
 These systematic differences are likely due  to the too low spatial resolution  of the
3D simulations  which are used here  (see Sect.~\ref{Influence of the resolution} below).

At  high frequency, the use of $\chi_k^{\rm E}$ instead of  $\chi_k^{\rm G}$ results in larger  ${\cal P}$
for all  stars. This arises from the fact that  $\chi_k^{\rm E}$
  spreads slightly more energy at   high frequency than  $\chi_k^{\rm
    G}$ does (see Fig.\ \ref{fig:kwpower}). 

The largest difference between the two types of  calculation (direct
versus semi-analytical) is seen  in the case of
Procyon. Indeed, the simulation of Procyon shows a pronounced
depression around $\nu \sim 1.5$\,mHz. Such a depression is not seen in the semi-analytical calculations.
The origin of this depression has not
been clearly identified yet but is perhaps related to some interference
 between the turbulence and the acoustic waves
 which manifests itself in  the pressure fluctuations in the 3D work integral but is
 not included in the semi-analytical description.


\subsection{Influence of the 3D simulation characteristics}
\label{Influence of the resolution}

  \begin{figure}[ht]
       \begin{center}
         \includegraphics[width=9.truecm]{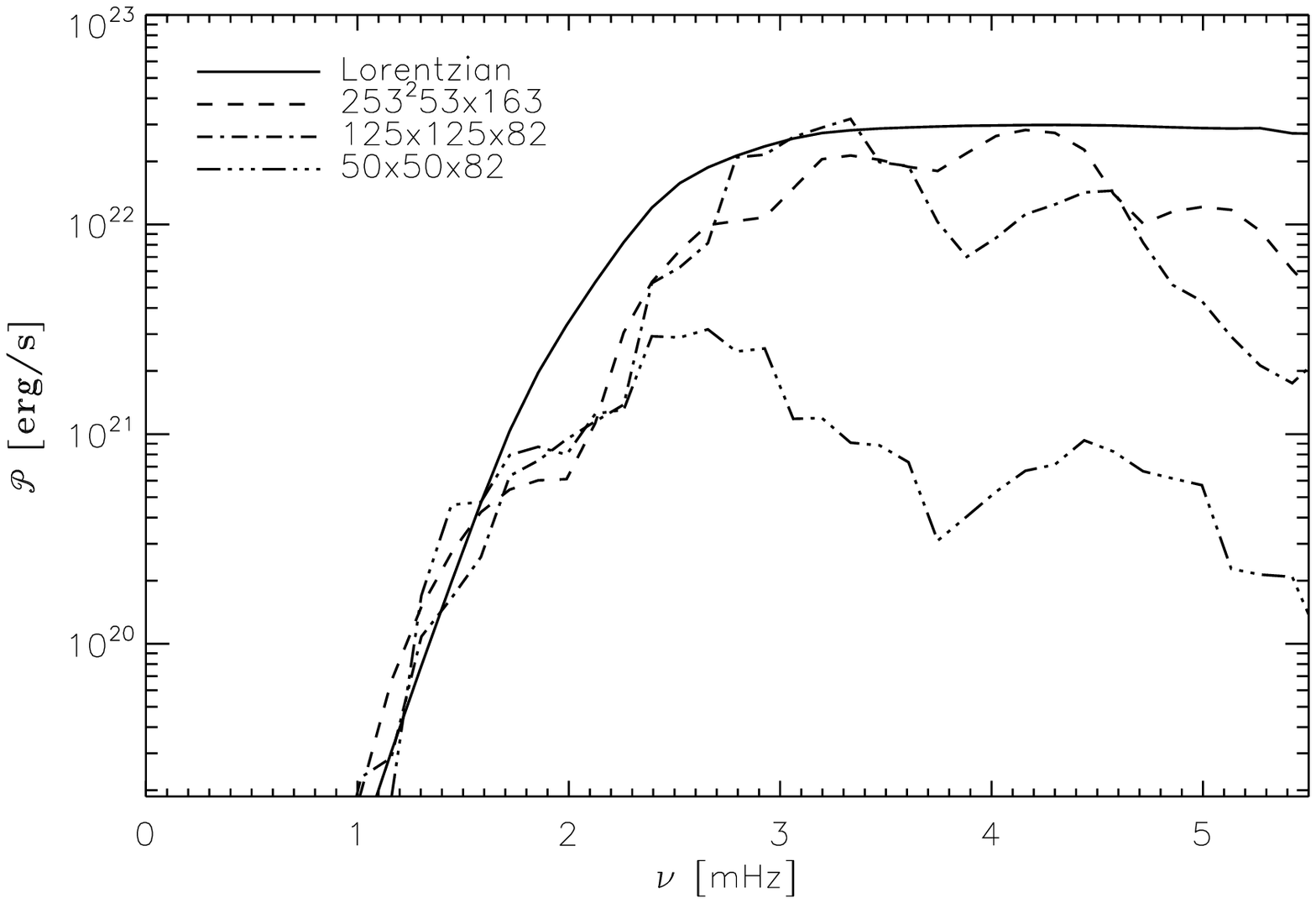} 
          \includegraphics[width=9.truecm]{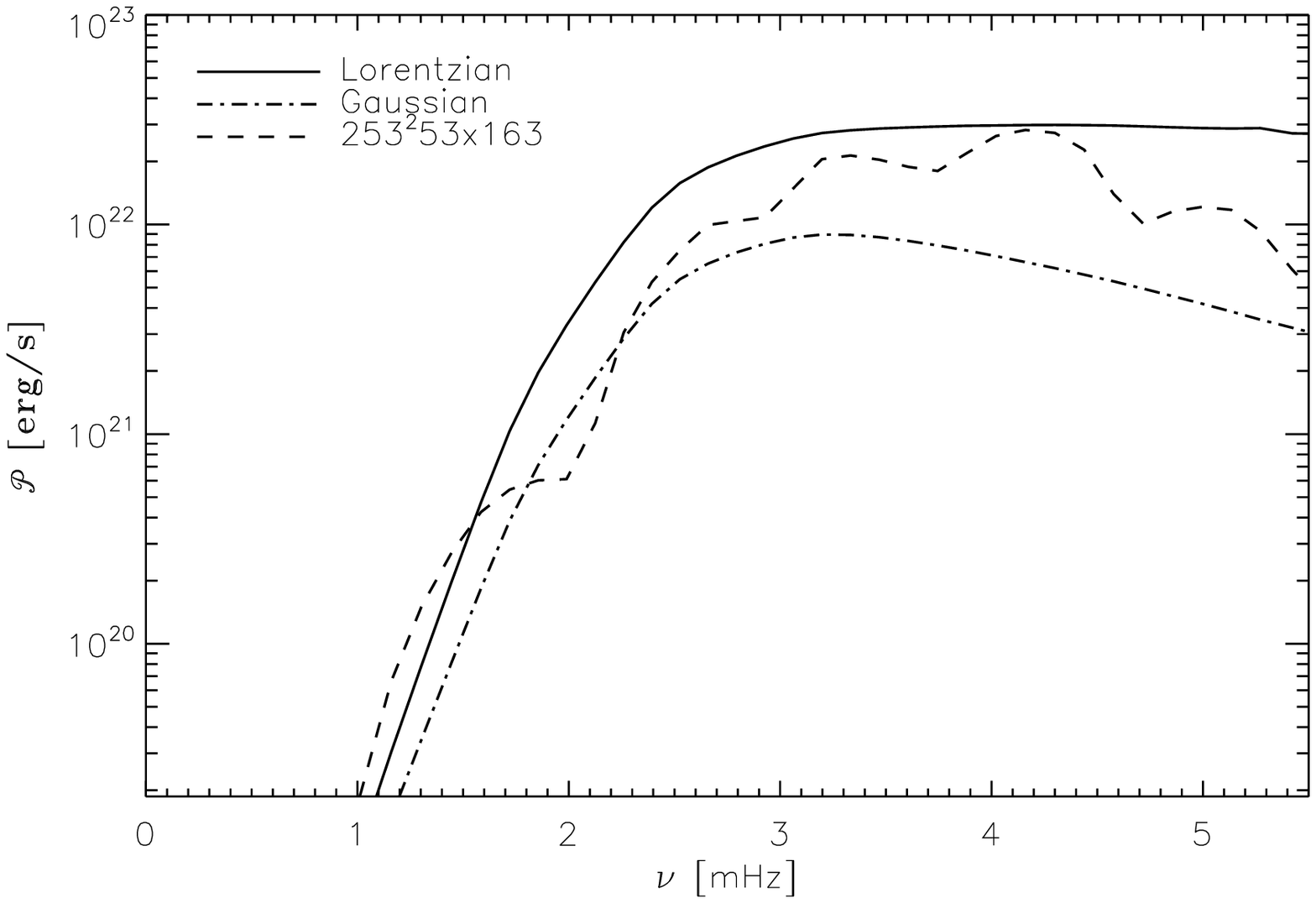} 
       \end{center}
\caption{Top: As in Fig.\,\ref{fig:P} for solar
   simulations only.  The solid line  corresponds to the semi-analytical calculations based on
a Lorentzian $\chi_k$ and a simulation with a spatial resolution of
   253$\times$253$\times$82.
The other lines are  running means over five frequencies
   of  the direct  calculation based on solar simulations with
   different spatial resolution: 253$\times$253$\times$82 (dashed line),
   125$\times$125$\times$82 (dot dashed line) and
   50$\times$50$\times$82 (dot dot dashed line) .
{Bottom}: The solid and dashed lines have  the same meaning as in
   the top figure. The dot-dashed line corresponds to the semi-analytical
   calculations based on a Gaussian  $\chi_k$.
}
\label{fig:Psun}
   \end{figure}

In order to assess the influence of the spatial resolution of the
 simulation on our results, we have at our disposal three other solar 3D
simulations, with a grid of $253
\times 253 \times 163$, $125 \times 125 \times 82$ and $50 \times 50
 \times 82$ (hereafter S1), and a duration of $\sim$ 42 min, 70 min and 100 min,
respectively.

 We have computed the  p-modes excitation rates according to the direct
 method for those three simulations.
 For each of those simulations we have also computed the
 excitation rates according to the semi-analytical method assuming
 either a Lorentzian $\chi_k$ or a Gaussian $\chi_k$.

As shown in
Fig.\,\ref{fig:Psun} (top), the excitation rates computed according to
the direct  calculation increase as the spatial resolution of the 3D
simulation increases. The excitation rates  computed with   the 3D simulations
with  the two highest spatial resolutions reach approximately the same mean
amplitude level, indicating that  this level of spatial resolution
is sufficient for the direct calculations.

 We note that as the spatial resolution increases, the semi-analytical calculations using a Lorentzian
$\chi_k$ decrease by a factor $\sim 2$ (not shown here).  The differences in
the semi-analytical calculations based on the  $253 \times 253 \times 163$ simulation and the $125
\times 125 \times 82$ simulation are found very small,  indicating that  this level of   spatial resolution
is sufficient for the semi-analytical calculations too.

Finally, we note that the excitation rates obtained  for the   $50 \times 50 \times 82$
 solar simulation (S1) are approximatively
 two times smaller than excitation rates for 
 the $50 \times 50 \times 82$ solar simulation
 otherwise used throughout this work (S0 hereafter).
 This difference is attributed to the fact that the two simulations
 do not correspond to the same realization.
 Indeed, as a test, we have extended the duration of the simulation S1 up to 500\
 min.
 The full time series has then been  divided into subsets   of equal
 duration of 100\ min  and p-mode excitation rates have been computed for each subset.
  We find that the  maximum in  the p-modes excitation rates ${\cal P}
 (\nu)$ oscillates from a subset to another about a mean value. The observed
 variations are large: the maximum in ${\cal P}
 (\nu)$ can be larger (smaller resp.) by $\sim$ 1.5 (0.5 resp.) times
 the  maximum in the power spectrum obtained by
 averaging the power spectra of all subsets.
  Hence we find that at low spatial resolution, different realizations yield
  excitation rates that are scattered about a mean value  at each frequency. This dispersion is
  likely to be responsible for the factor of two difference between  the excitation rate maxima
  obtained for the two realizations S0 and S1.
   This type of dependency of  ${\cal P}$ ---  with the starting time of the time
 series and its  duration ---  is expected to be smaller for simulations
 with resolution higher than $50 \times 50 \times 82$, because of the larger
 number of excitation sources there. This will be studied in a subsequent work.

\subsection{Eddy-time correlation: Lorentzian versus Gaussian}

As seen in Sect~\ref{Influence of the resolution} above, 
 the characteristics of the simulations influence  the semi-analytical
calculations of the mode excitation rates (through the input parameters which
enter the semi-analytical calculations and which are taken
from the 3D simulation).  We want to compare the results
of the semi-analytical calculations using  $\chi_k^{\rm L}$ 
with the  semi-analytical calculations using $\chi_k^{\rm G}$. It is
then  necessary to insert the 3D inputs in these calculations  coming 
from simulations with the highest quality, here the highest
available resolution.

Fig.\,\ref{fig:Psun} (bottom)  compares semi-analytical calculations
using  a Lorentzian  $\chi_k$ with those using a Gaussian
$\chi_k$. All theses semi-analytical calculations are here based on
the energy spectrum of the simulation with
the spatial resolution of $253 \times 253 \times 163$ (see Sect.~\ref{Influence
  of the resolution}). 

The average level of the excitation rates calculated according to the
 direct method and with  the simulation with the highest spatial
 resolution  is in between the semi-analytical calculations based on Lorentzian
 $\chi_k$ and those based on a Gaussian $\chi_k$, nevertheless they
 are  in general slightly closer to the semi-analytical calculations based on Lorentzian
 $\chi_k$. 
This result is discussed in Sect.~\ref{sec:The eddy time-correlation:Solar case}.

\subsection{Maximum of ${\cal P}$ as a function of $L/M$}
\label{sec:pmax}

Fig.\ \ref{fig:pmax} shows  ${\cal P}_{\rm max}$, the maximum in ${\cal P}$,
as a function of $L/M$ for the direct  and the  semi-analytical  calculations.

The same  systematic differences between the direct and the
semi-analytical calculations  as seen in Fig.\ \ref{fig:P} 
are of course observed here. Note that the differences slightly decrease  with increasing values of $L/M$.

We have also computed the excitation rate with the semi-analytical method using  $\chi_k^{\rm L}$ . 
The maximum excitation rate as evaluated with $\chi_k^{\rm L}$ is systematically larger than both the direct calculations 
and the semi-analytical results based on $\chi_k^{\rm G}$ or $\chi_k^{\rm E}$.

 In the solar case, ${\cal P}_{\rm max}$ is found to be closer to the value 
derived from  recent helioseismic data \citep{Baudin05} when using a
Lorentzian compared to a Gaussian \citep[see also][B06b hereafter]{Belkacem06b}.
The 'observed' excitation rates are derived from the velocity observations $V$ as follows:
\eqn{
{\cal P}= 2 \pi ~ \Gamma_\nu ~ \mathcal{M}(h) ~ V^2
\label{eqn:P:V}
}
where $\mathcal{M}$ is the  mode mass, 
$V$ is the  mode velocity amplitude and $h$ is the height above the
photosphere where the mode mass is evaluated. The mode line width at half maximum in Hz, $\Gamma_\nu = \eta/ \pi $,
 ($\eta$ is the mode amplitude damping rate in $s^{-1}$)  is determined observationally in the solar case.\\
Using the  recent helioseismic measurements of $V$ and $\Gamma_\nu$ by \citet{Baudin05} and the  mode mass computed
here for our solar model  at the height $h$=340\ km \citep[cf.][]{Baudin05}, we find ${\cal
  P}_{\rm max,\odot}= 6.5 \pm 0.7 ~\times 10^{22}$\,erg\,s$^{-1}$. 
This value must be compared with those found with $\chi_k^{\rm L}$ and
$\chi_k^{\rm G}$, namely 
 ${\cal P}_{{\rm max},\odot}^{\rm L}=  4.9 \times
10^{22}$\,erg\,s$^{-1}$ and   ${\cal P}_{{\rm max},\odot}^{\rm G}=
1.2\times 10^{22}$\,erg\,s$^{-1}$ respectively. 
%

\bigskip
{\it Scaling laws:} All sets of calculations can be  reasonably well fitted with a scaling law of the
 form  ${\cal P}_{\rm max} \propto (L/M)^{s}$ where $s$ is a slope which
depends on the considered set of calculations.  Values found for $s$ are summarized in Table~\ref{tab:slopes}.

$\bullet$ For the semi-analytical calculations, we find $s=2.6$ using $\chi_k^{\rm L}$,
$s=3.0$ using $\chi_k^{\rm E}$ and  $s=3.1$ for the Gaussian form.

The Lorentzian form results in a power law with a smaller slope than the Gaussian. This  can be understood as follows:
A Gaussian decreases more rapidly with $\nu$ than a Lorentzian.
As the ratio $L/M$  of a main sequence star  increases, the mode frequencies shift to lower values.
Hence p-modes of stars with large values of  $L/M$ receive \emph{relatively} more acoustic energy
when adopting a Gaussian rather than a Lorentzian $\chi_k$. It is worthwhile to note that even though the ratio $L/M$ is the ratio
 of two global stellar quantities, it nevertheless characterizes essentially the stellar surface layers where the mode excitation is located since $L/M \propto T_{\rm eff}^4/g$.

$\bullet$ For the  set of direct calculations,  some scatter exists
as a consequence of the large statistical fluctuations in ${\cal P}_{\rm max}$ and
a linear regression gives   $s=3.4$. As expected,  this value is rather close to that found with the semi-analytical
  calculations using  either $\chi_k^{\rm G}$ or $\chi_k^{\rm E}$.

\fig{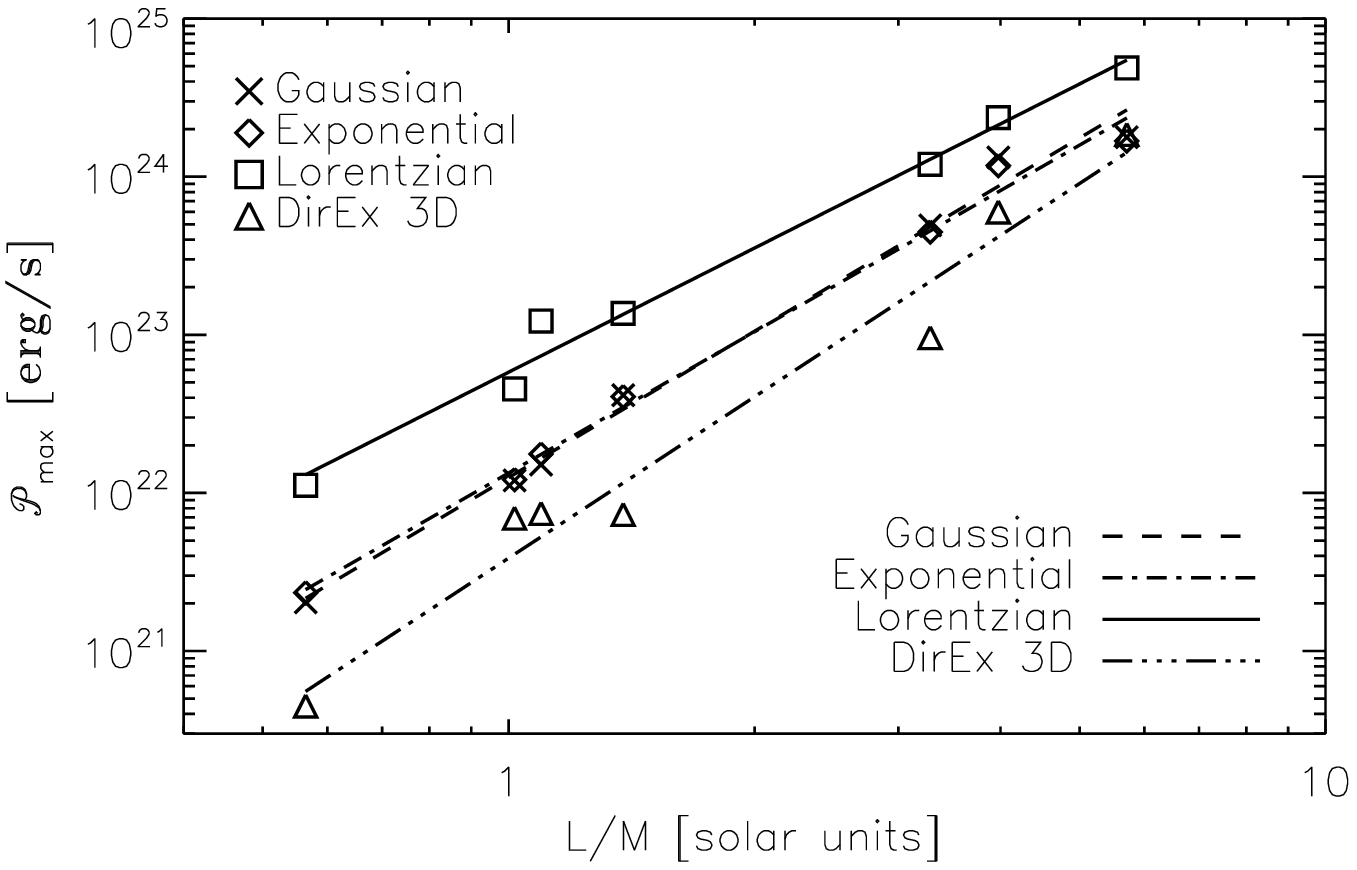}{${\cal P}_{\rm max}$ versus $L/M$ where $L$ is the
luminosity and $M$ is the mass of the 1D~models associated with the 3D simulations.
The triangles correspond to the  direct calculations (labeled as 'DirEx 3D' in the legend),
and the other symbols correspond to the semi-analytical calculations using the three forms of $\chi_k$:
the crosses assume a Gaussian, the diamonds an exponential and the squares a Lorentzian, respectively.
Each set of ${\cal P}_{\rm max}$ is fitted by a  power law of the
form $(L/M)^{s}$ where $s$ is the slope of the power law.
The line-styles correspond to the three semi-analytical cases and the
direct calculations, as indicated in the lower right corner of the plot.
}{fig:pmax}

\begin{table}[ht]
\begin{center}
\begin{tabular}{cccc}
method &    $\chi_k$ & $s$  & $sv$   \\
\hline
direct          &  ---         & 3.4   & --- \\
semi-analytical &  Gaussian    & 3.1 &  1.0\\
semi-analytical &  exponential & 3.0 &  0.9\\
semi-analytical &  Lorentzian  & 2.6 &  0.7\\
\end{tabular}
\end{center}
\caption{Values found for the slopes $s$ (see Sect.\ \ref{sec:pmax})
and $sv$ (see Sect.\ \ref{sec:vmax}). 'Method' is the method considered for the calculations of ${\cal P}$.  }
\label{tab:slopes}
\end{table}

\subsection{Maximum of the mode amplitudes ($V_{\rm max}$) as a function of $L/M$}
\label{sec:vmax}

The theoretical oscillation velocity  amplitudes $V$ can be computed 
according to Eq.\ (\ref{eqn:P:V}) 
The calculation requires the knowledge of  the excitation rates, ${\cal P}$, damping rates, $\eta$, and mode mass, ${\cal M}$. 
Although it is possible --\,in principle\,--
to compute the convective dampings from the 3D simulations \citep{Stein01I},
it is a difficult task which is under progress.
However, using for instance Gough's Mixing-Length Theory \citeyearpar[][ G'MLT hereafter]{Gough76,Gough77},
it is possible to compute $\eta$ and ${\cal P}$ for different stellar models of given $L,M$ and deduce
  $V_{\rm max}$, the maximum  of the mode amplitudes,  as a function of $L/M$ at the cost of
some inconsistencies.

In \citet{Samadi01b}, calculations  of the damping rates $\eta$ based on G'MLT were performed
for  stellar models with different values of  $L$ and $M$.
Although these stellar models are not the same as those  considered here,
it is still  possible, for a crude estimate, to determine the  dependency of $V_{\rm max}$
with $L/M$.

Hence  we proceed as follows:
For each stellar model computed in \citet{Samadi01b},
 we derive the values of  $\eta$  and  $\mathcal{M}$ at  the frequency $\nu_{\rm max}$
at which the maximum amplitude is expected.
From the stars for which solar-like oscillations have been detected,
\citet{Bedding03} have shown that this frequency is proportional to the cut-off frequency.
Hence we determine $\nu_{\rm max}=(\nu_{\rm c}/ \nu_{{\rm c},\odot}) \, \nu_{\rm max,\odot} $
where $\nu_{\rm c}$ is the cut-off frequency of a given model
and the symbol $\odot$  refers to solar quantities
($\nu_{\rm max,\odot} \simeq 3.2$\,mHz and $\nu_{c,\odot} \simeq 5.5$\,mHz).
We then obtain ($\eta_{\rm max} ~ \mathcal{M}_{\rm max}$) as a function of $L$ and $M$.

On the other hand, in Sect.\ \ref{sec:pmax}, we have  established ${\cal P}_{\rm max}$ as a function of $L$ and $M$.
Then, according to Eq.\ (\ref{eqn:P:V}), we can determine $V_{\rm max}(L,M)$ for the different power laws of  ${\cal P}_{\rm max}$.

 We are interested here in the slope (i.e. variation with $L/M$) of $V_{\rm max}$ and 
not its absolute magnitude, therefore we scale the theoretical and observed $ V_{\rm max}$
with a same normalization value which is  taken as the solar value 
  $V_{\rm max,\odot}= 33.1\,\pm 0.9$\,cm\,s$^{-1}$
as determined  recently by \citet{Baudin05}.
 

We  find that  $V_{\max}$ increases as $(L/M)^{sv}$ with  different values for $sv$
depending on the assumptions for $\chi_k$.
The values of $sv$ are summarized in Table~\ref{tab:slopes}
and illustrated in Fig.\ \ref{fig:vmax}. We find $sv\simeq 0.7$
with $\chi_k^{\rm L}$ and  $sv \simeq 1.0$ with $\chi_k^{\rm G}$. 

These scaling laws must be compared with 
observations of a few stars for which solar-like oscillations have been detected in Doppler velocity.
The observed $V_{\rm max}$ are taken from Table 1 of HG02, except
for $\eta$\,Boo, $\zeta$~Her~A, $\beta$~Vir, HD~49933 and $\mu$~Ara, for which we use the $V_{\rm max}$ quoted by \cite{Carrier02}, 
\citet{Martic01}, \citet{Martic04}, \citet{Mosser05} and \citet{Bouchy05} respectively and $\epsilon$~Oph and $\eta$~Ser quoted by \citet{Barban04}.

Fig.\ (\ref{fig:vmax})  shows that the observations also indicate a monotonic logarithmic increase of $V_{\rm max}$  with $L/M$  despite 
 a  large dispersion  which may  at least partly arise  from different origins of the data sets. 
For the observations we find a 'slope' $sv \simeq 0.7$.
This is  close to the theoretical slope obtained  when adopting  $\chi_k^{\rm L}$
and definitely lower than the slopes obtained when adopting   $\chi_k^{\rm G}$ or adopted by HG02.


%
%
\fig{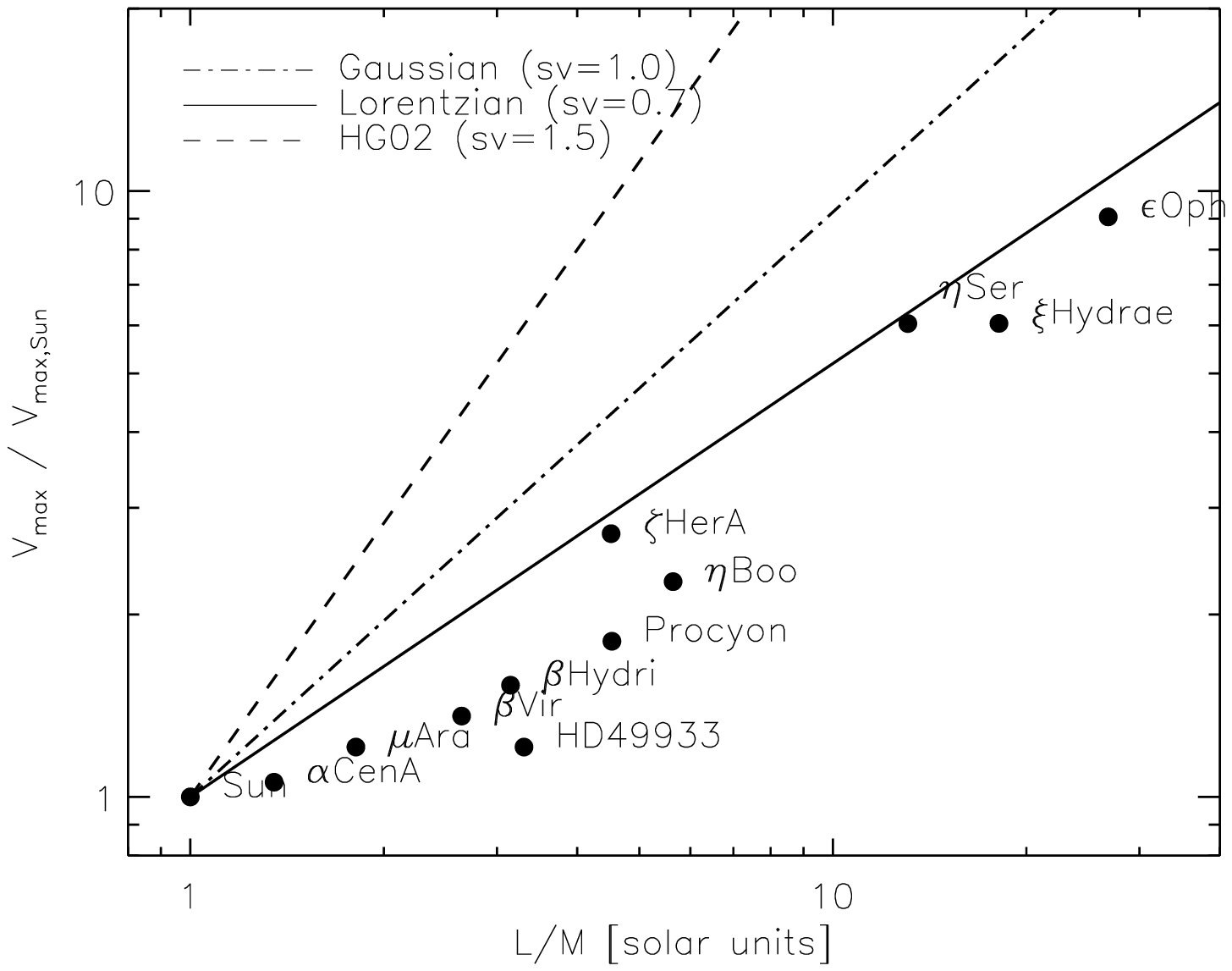}{Same as Fig.\ \ref{fig:pmax}  for $V_{\rm max}/V_{{\rm max},\odot}$,
the maximum of the mode amplitudes relative to the observed solar
value ($V_{{\rm max},\odot}=33.1\,\pm 0.9$\,cm\,s$^{-1}$).
 The filled  symbols correspond to the stars
for which solar-like oscillations  have been  detected in Doppler velocity.
The lines - \ except the dashed line \ -  correspond to the power laws 
obtained from the predicted scaling laws for  ${\cal P}_{\rm max}$
(Fig.\ \ref{fig:pmax}) and estimated values of the 
damping rates $\eta_{\rm max}$ (see text for details).
Results for two different eddy time-correlation functions, $\chi_k$,
are presented:  Lorentzian (solid line) and Gaussian (dot-dashed line) functions. 
For comparison the dashed line shows the result by HG02. Values of the
slope $sv$ are given on the plot and in Table~3.}{fig:vmax}

\section{Summary  and discussion}
\label{sec:conclusion}

 One goal of the present work  has been
 to  validate the model of stochastic excitation presented in Paper\,I.
The result of this test is summarized in Sect.\ \ref{sec:Validation of the MSE}.
A second  goal has been to study the properties of  the turbulent eddy time-correlation, $\chi_k$,
and the importance for the calculation of the  excitation rates, ${\cal P}$, of the adopted form of $\chi_k$.
Section \ref{sec:The eddy time-correlation} deals with this subject.

\subsection{Validation of the excitation model}
\label{sec:Validation of the MSE}

In order to check the validity of the excitation model, seven 3D simulations of stars, including  the Sun,
have been considered.
For each simulation, we calculated the p-mode excitation rates, ${\cal P}$,
using two methods: the semi-analytical excitation model
({\latin{cf.\ }}Sect.\ \ref{sec:The semi-analytical method}) that we are testing,
and a direct calculation  as detailed in
Sect.\ \ref{sec:Calculation of the p mode excitation rates}.
In the latter method, the work performed by the pressure fluctuations
on the p-modes is calculated directly from the 3D simulations.

In the  semi-analytical method, ${\cal P}$ is computed according  to the  excitation model of Paper\,I. The calculation  uses,
as input, information from the 3D simulations as for instance
the eddy time-correlation ($\chi_k$) and the kinetic energy spectra ($E(k)$).
However although $\chi_k$ has been   computed for each simulation, in practice for simplifying the
problem of implementation as well as for comparison purpose with Paper\,III,
we chose to represent the $\nu$ variation of   $\chi_k$ with simple analytical functions.
It is  found that  the $\nu$-variation of $\chi_k$ in the present simulations  lies
loosely  between that of an exponential and a Gaussian.  We then perform the validation test of the excitation model
using those two forms of $\chi_k$.

\bigskip

 We find that using either  $\chi_k^{\rm G}$ or $\chi_k^{\rm E}$ in
 the semi-analytical calculations of  ${\cal P}$ results in
 systematically higher excitation 
rates than those obtained with direct 3D calculations.
 These  systematic differences are attributed to the low spatial resolution of  our present set
  of simulations. Indeed we have shown here that  using  solar
 simulations with different spatial resolutions, the resulting excitation rates increase
  with increasing spatial resolution.

 We have next investigated the dependence of   ${\cal P}_{\rm max}$ with $L/M$
(See Fig.\ \ref{fig:pmax}),
where $L$ and $M$ are the stellar luminosity and mass respectively.
As in previous works based on a purely theoretical approach \citep[{\protect{\latin{e.g.\ }}}][]{Samadi02b},
 we find that ${\cal P}_{\rm max}$  scales approximatively as $(L/M)^{s}$ where $s$ is
the slope of the scaling law: we find $s=3.4$ with the  direct
calculations and $s=3.2$ and $s=3.1$ with the  semi-analytical
calculations using $\chi_k^{\rm G}$ and  $\chi_k^{\rm E}$
respectively. This indicates a general agreement between the
scaling properties of both types of
calculations, 
which validates to some extent the adopted excitation model
across the  domain of the HR diagram studied here.

For the sake of simplicity, only simple analytical
 forms for $\chi_k$ have been investigated here.
 We  expect that the  use of more sophisticated forms for $\chi_k$
would reduce the dispersion between the analytical and direct calculations,
but would not affect the conclusions of the present paper.

\subsection{The eddy time-correlation spectra, $\chi_k$ }
\label{sec:The eddy time-correlation}



The slope $s$  of the scaling law for ${\cal P}_{\rm max}$, is found to
depend significantly on the adopted analytical form for $\chi_k$.
 The semi-analytical calculations using the Lorentzian form
  for  $\chi_k$ results in a significantly smaller slope $s$ than those based on the
  Gaussian or the exponential or from direct calculations (see Table\ \ref{tab:slopes}).

Except for the Sun, independent and accurate enough constraints on \emph{both} the
mode damping rates and the mode excitation rates are not yet
available. We are then left to perform comparison between predicted and
observed mode amplitudes. Unfortunately,  obtaining tight constraints
on $\chi_k$ using comparison between predicted and observed mode amplitudes 
is hampered by large uncertainties in the theoretical estimates of the damping rates. 
It is therefore currently difficult to derive the excitation rates ${\cal P}$
for the few stars for which solar-like oscillations have been detected \citep[see][]{Samadi04b}.
The future space mission COROT \citep{Baglin98} will provide high-quality data on seismic
observations. Indeed the COROT mission will be the first mission that will provide
\emph{both} high precision mode amplitudes and line-widths for stars other than the Sun.
It will then be possible to use  the observed damping rates and   to
derive the excitation  rate ${\cal P}$ free of the uncertainties associated with a theoretical computation of damping rates.
In particular, it will be possible  to determine  ${\cal P}_{\rm max}$
as a function of $L$ and $M$ from the observed  stars.
Such observations will provide valuable constraints for our models for $\chi_k$.

We can, nevertheless, already give  some arguments below in favor
 of  the Lorentzian being  the correct  description for $\chi_k$.


\subsubsection{Solar case}
\label{sec:The eddy time-correlation:Solar case}

In the 3D simulations studied here, including that of the Sun,
the inferred  $\nu$ dependency of $\chi_k$ is far from a Lorentzian, in contrast
to that found with the solar 3D simulation investigated in Paper\,III.
However, by investigating solar simulations with different resolutions,
we find that, as  the spatial resolution increases,  $\chi_k$  tends towards
a Lorentzian $\nu$-dependency.
This explanation is  likely to stand for non-solar simulations too,
but has not yet been confirmed (work in progress).

 Furthermore, as shown in Fig.\
\ref{fig:Psun}, bottom, the direct calculations obtained with the simulation with 
the highest  spatial resolution available is slightly closer to the  semi-analytical
calculations using the Lorentzian form than those using  the Gaussian
 one.

Independently of the resolution (if large enough of course),  a
Lorentzian $\chi_k$ predicts larger values for ${\cal P}_{\rm max}$
than a Gaussian or an exponential do. 
In particular in the solar case,  the semi-analytical calculation  using
$\chi_k^{\rm L}$ results in a ${\cal P}_{\rm max}$ closer to the
helioseismic constraints  derived by \citet{Baudin05}
compared to using $\chi_k^{\rm G}$ or $\chi_k^{\rm E}$.
This latter result is in agreement with that of  Paper\,III.

Part of the remaining discrepancies with the helioseismic constraints are
attributed to the adopted closure model according to \citet[][B06b hereafter]{Belkacem06b}. 
Indeed, theoretical models of stochastic excitation adopt the
quasi-normal approximation (QNA). As shown in B06b,  the skew introduced by the QNA
 result in a under-estimation of the solar p mode excitation rates.  
 When the so-called closure model with plumes proposed by 
 \citet{Belkacem06a} is adopted, new semi-theoretical calculations fit
 rather well  the recent  helioseismic constraints derived by
 \citet[][see B06b]{Baudin05}.

\subsubsection{$V_{\rm max}$ as a function of $L/M$}

Consequences of the predicted power laws for  ${\cal P}_{\rm max}$
have also been crudely investigated here for the expected value of $V_{\rm max}$,
the maximum value of the mode velocity (Fig.\ \ref{fig:vmax}).
Calculations of  $V_{\rm max}$ from  ${\cal P}_{\rm max}$ require the  knowledge
of the mode damping rates, $\eta$, which cannot be fully determined from the simulations.
We are  then led to use theoretical calculations of the damping rates.
We consider here  those  performed by \citet{Samadi01b}
which are based on Gough's \citeyearpar{Gough76,Gough77}
non-local and time-dependent formulation of convection.
From those values of $\eta$ and  the  different power laws for  ${\cal P}_{\rm max}$
 expected  values of $V_{\rm max}$ are obtained.

We find, as in \citet{Houdek02} (HG02), that $V_{\rm max}$ scales as $(L/M)^{sv}$.
Calculations by HG02 result in  $sv\simeq 1.5$. 
Our semi-analytical calculations of  ${\cal P}_{\rm max}$  based on a Gaussian $\chi_k$ result in a slightly smaller slope ($sv\simeq 1.0$).
On the other hand,  using a Lorentzian $\chi_k$ results in a slope
$sv\simeq 0.7$ 
which   is closer to that derived from the few stars for which  oscillation
amplitudes have been measured.

From this result, we  conclude that the problem of the \emph{over}-estimation of
the amplitudes of the solar-like oscillating  stars  more luminous than the Sun is
 related to  the choice of the model for $\chi_k$.
Indeed, previous theoretical calculations by \citet{Houdek99} are based on the assumption of a Gaussian $\chi_k$.
 As shown here, the Gaussian assumption results in a larger slope $sv$
than the Lorentzian $\chi_k$. This is the reason why \citet{Houdek99} \emph{over}-estimate $V_{\rm max}$ for
$L/M > L_\odot/M_\odot$. 
\\
On the other hand, if one assumes $\chi_k=\chi_k^{\rm L}$, a scaling factor is no longer required
 to reproduce  ${\cal P}_{\rm max}$ for the solar p-modes. Moreover,
as a consequence of the smaller slope, $sv$, resulting from a Lorentzian $\chi_k$,
the predicted amplitudes for other stars match the observations better.

This result further indicates that a Lorentzian is the better choice for
$\chi_k$, as was also concluded in Paper\,III.

Departures of the theoretical curve from the observed points in
Fig.\ \ref{fig:vmax} can be attributed to several
causes which remain to be investigated:

\begin{itemize}
\item[1)] A major uncertainty comes from the computed damping rates
as no accurate enough observations are available yet to validate them.
As $V$ results from the balance  between  ${\cal P}$ and  $\eta$, the slope $sv$ can also
depend on the variation of  $\eta$ with $L/M$.
Thus, the large differences in $sv$ between the seismic observations and the  calculations based on
$\chi_k^{\rm G}$ can also be, {\latin{a priori}}, attributed to an incorrect evaluation of the damping rates.
However   $\eta_{\max}$ -\ the value of the damping rate at the frequency  $\nu_{\rm max}$  at which
 the maximum amplitude is expected \ - does not follow a clear scaling law  with $L/M$. 
We have looked at the $\eta_{\max}$ variation in our set of G'MLT models
and found no clear dependence of $\eta_{\max}$ on $L/M$ but rather a dispersion.

\item[2)] The observed stars in Fig.\ \ref{fig:vmax} have  somewhat different chemical compositions;
this can cause some scatter in the relation $V_{\rm max}$-$L/M$ which has  not  been taken into account here.
 All the simulations investigated in the present work employ a solar metal abundance.
The  metallicity  has a direct impact on the opacity and the EOS.
Both in turn affect the internal structure and are also decisive for the
transition from convection to radiation in the photosphere and therefore
determine the structure of the super-adiabatic region.
Hence,  the properties of the super-adiabatic
region,  relevant for the excitation rates, differ for
stars located  at the same position in the HR diagram ({\latin{e.g.,}} same $T_{\rm eff}$ and same $g$)
but with different metal abundances.
Consequently  the excitation of p-modes for such stars  probably differ,
although it remains to be seen to what extent.
A differential investigation of the metallicity effect is planned for the
future.
\end{itemize}

\subsection{Relative contribution of the turbulent pressure}

Another issue concerns the {\it relative contribution of the turbulent pressure.}
The excitation of solar-like oscillations  is generally attributed to the turbulent pressure  ({\latin{i.e.}} Reynolds stress) and
the entropy fluctuations ({\latin{i.e.}} non-adiabatic gas pressure fluctuations)
and  occurs in the super-adiabatic region where those two terms  are the largest.
In Paper\,III, it was found that the two driving sources are of the same order of magnitude,
in contradiction with the results by \citet{Stein01II}
who found -- based on their 3D numerical simulations of convection --
that the turbulent pressure is the dominant contribution to the excitation.
The discrepancy  is removed here as we used a corrected version of
the formulation of the contribution of the Reynolds stress of Paper\,I (see Eq.~(\ref{eqn:SR})), leading to a larger contribution from the Reynolds stress.



For the Sun, assuming $\chi_k^{\rm L}$ ($\chi_k^{\rm G}$ resp.),
we now find that
the Reynolds stress contribution is 5~times (3~times  resp.) larger
than that  due to the entropy fluctuations (non-adiabatic gas-pressure fluctuations).
Hence,  the Reynolds stress is indeed the dominant source of excitation
in agreement with the results of \citet{Stein01II}. The best agreement with the latter results is obtained with a Lorentzian $\chi_k$.

However, we find that the relative contribution from Reynolds stresses decreases rapidly with  $(L/M)$.
For instance, in the simulation of Procyon, the Reynolds stress
represents only $\sim 30\,\%$ of the total excitation rate.

From that, we conclude  that the excitation by entropy fluctuations  cannot
be neglected, especially for stars more luminous than the Sun.

\begin{acknowledgements}
RS's work has been supported by  Soci\'et\'e de Secours des Amis des Sciences (Paris, France)
and by  Fundac\~ao para a Ci\^encia e a Tecnologia (Portugal) under grant SFRH/BPD/11663/2002. 
RFS is supported by NASA grants NAG 5 12450 and NNG04GB92G and by NSF grant AST0205500.
We thank the referee, Mathias Steffen, for his judicious suggestions which helped to improve this manuscript.
\end{acknowledgements}


\appendix
\section{ Calculation of the non adiabatic pressure fluctuations}

The adiabatic variation of the gas pressure does not contribute to the
$\Delta (P dV)$  work over an oscillation period as it is in phase with the volume (or density) variation.
In practice, however, it is beneficial for
the accuracy of the computation of excitation to
subtract the adiabatic part of the gas pressure fluctuation, since it
reduces the coherent part.  That part gives zero contribution only in
the limit of infinite time, or for an exact integer number
of periods. However, in practice, it gives rise to a random (or noisy)
contribution. Indeed, as we deal with a lot of different modes
 it is  hard to find a time-interval which is an integer
number of periods of each and all of the modes at the same time.

The Lagrangian variations of gas pressure, $\Delta P_{\rm gas}$ must  satisfy 
\begin{equation} {\Delta P_{\rm gas} } = { \Gamma_1  P_{\rm gas}
    \over \rho }\; \Delta \rho   + {\partial P_{\rm gas}\over \partial
    S} \Delta S
\label{dpnadgas_2}
\end{equation}
where  $P_{\rm gas}$, $\rho$ and $S$ are the
gas pressure the density and the entropy respectively 
and where the operator $\Delta $ represents the \emph{pseudo} Lagrangian
fluctuations of a given quantity. 
The concept of pseudo  Lagrangian fluctuations is introduced in
Nordlund \& Stein (2001).
Accordingly we  derive the non-adiabatic gas pressure fluctuations as: 
\eqna{
{\Delta P }_{\rm gas,nad}(\vec r,t) & \equiv & {\Delta P_{\rm gas} } - c_s^2 \; \Delta \rho  } 
where $c_s^2 \equiv \Gamma_1 P_{\rm gas}  / \rho $ is the sound speed.

However, what we want to
subtract off from $\Delta P_{\rm gas}$
is that part of the pressure variation that is due
to adiabatic compression and expansion due to the particular \emph{radial
wave modes}  (i.e. the low amplitude perturbation of $\rho(r)$ on top
of the possibly large variations horizontally of $\rho(\vec r)$  that
$\rho(r) $ is an average of). 

To find the nonadiabatic pressure fluctuations, 
we start with calculations of horizontal averages of the primary quantities, 
$P_{\rm gas}$, $P_{\rm turb}$, $\rho$ and $c_s^2$. We convert these averages 
to the pseudo-Lagrangian frame of reference, in which the net mass flux vanishes.
We then compute fluctuations of the resulting quantities
with respect to time, {\rm i.e.}, subtract their time averages:
\eqna{
{\Delta P}_{\rm gas} = \langle P_{\rm gas} \rangle_{\rm h} - 
\langle P_{\rm gas} \rangle_{\rm h, t} \nonumber \\
{\Delta P}_{\rm turb} = \langle P_{\rm turb} \rangle_{\rm h} -
\langle P_{\rm turb} \rangle_{\rm h, t} \nonumber \\
{\Delta \rho} = \langle \rho \rangle_{\rm h} -
\langle \rho \rangle_{\rm h, t} \nonumber
}
Here, $\langle \rangle_{\rm h}$ refers to horizontal average and 
$\langle \rangle_{\rm h,t}$ refers to consequent time average performed on
a horizontally averaged quantity.
Finally, the non-adiabatic fluctuations of the \emph{total} pressure 
(that is gas + turbulent pressure) are:
\eqna{
{\Delta P }_{\rm nad} = 
{\Delta P }_{\rm gas,nad} + {\Delta P }_{\rm turb} \nonumber \\
= {\Delta P} - \langle c_s^2 \rangle_{\rm h,t}  \; \Delta \rho
\label{dpnad_2}
}
where $\Delta P \equiv \Delta P_{\rm gas} + \Delta P_{\rm turb}$.

--------------------------------------------------------------------



\end{document}